\documentclass{appolb}

\usepackage{graphicx}
\usepackage{amsmath}
\usepackage{subfig}
\usepackage{cite}
\usepackage{color}

\newcommand{\msbar}{\ensuremath{\overline{MS}} }
\newcommand{\Ord}{\ensuremath{{\cal O}}}

\newcommand{\ZMVFNS}{\ensuremath{\text{ZM-VFNS}}}
\def\gsim{\mathrel{\rlap{\lower4pt\hbox{\hskip1pt$\sim$}} \raise1pt\hbox{$>$}}} 
\def\lsim{\mathrel{\rlap{\lower4pt\hbox{\hskip1pt$\sim$}} \raise1pt\hbox{$<$}}}  

\begin{document}
\title{
\vspace{-2.5cm}
\begin{flushright}
		{\small {\bf SMU-HEP-12-07}}   \\
\end{flushright}
\vspace{0.5cm}
Heavy Quark Production in the ACOT Scheme\\ beyond NLO%
%
\thanks{Presented by A.~Kusina at the
{\em Cracow Epiphany Conference 2011 - on Present and Future of B-Physics},
9-11, January, 2012.}%
}

\author{T.~Stavreva$^a$, F.~I.~Olness$^b$, I.~Schienbein$^a$, T.~Je\v{z}o$^a$,
A.~Kusina$^b$, K.~Kova\v{r}\'{\i}k$^c$, J.~Y.~Yu$^{a,b}$
\address{$^a$ Laboratoire de Physique Subatomique et de Cosmologie, Universit\'e
Joseph Fourier/CNRS-IN2P3/INPG,
 53 Avenue des Martyrs, 38026 Grenoble, France}
\address{$^b$ Southern Methodist University, Dallas, TX 75275, USA}
\address{$^c$ Institute for Theoretical Physics, Karlsruhe Institute of Technology, 
Karlsruhe, D-76128, Germany}
}

\maketitle
\begin{abstract}

We compute the structure functions $F_{2}$ and $F_{L}$ in the ACOT
scheme for heavy quark production.
We use the complete ACOT results to NLO, and make use of the $\msbar$
massless results at NNLO and N$^{3}$LO to estimate the higher order
mass-dependent corrections.
We show numerically that the dominant heavy quark mass effects can be
taken into account using massless Wilson coefficients together with an
appropriate rescaling prescription.
Combining the exact NLO ACOT scheme with these expressions should
provide a good approximation to the full calculation in the ACOT
scheme at NNLO and N$^{3}$LO.

\end{abstract}
\PACS{P12.38.-t,12.38Bx,12.39.St,13.60.-r,13.60.Hb}

\section{Introduction\label{sec:intro}}


The production of heavy quarks in high energy processes has become
an increasingly important subject of study both theoretically and
experimentally. The theory of heavy quark production in perturbative
Quantum Chromodynamics (pQCD) is more challenging than that of light
parton (jet) production because of the new physics issues brought
about by the additional heavy quark mass scale. The correct theory
must properly take into account the changing role of the heavy quark
over the full kinematic range of the relevant process from the threshold
region (where the quark behaves like a typical {}``heavy particle'')
to the asymptotic region (where the same quark behaves effectively
like a parton, similar to the well known light quarks $\{u,d,s\}$).

With the ever-increasing precision of experimental data and
the progression of theoretical calculations and parton distribution function (PDF) 
evolution to next-to-next-to-leading order (NNLO) of QCD there is a clear
need to formulate and also implement the heavy quark schemes at
this order and beyond.
The most important case is arguably the heavy quark treatment
in inclusive deep-inelastic scattering (DIS) since the very precise
HERA data for DIS structure functions and cross sections form the backbone
of any modern global analysis of PDFs. Here, the heavy quarks
contribute up to 30\% or 40\% to the structure functions at small momentum
fractions $x$.
Extending the heavy quark schemes to higher orders is therefore 
necessary for extracting precise PDFs, and this is a prerequisite for precise predictions
of observables at the LHC.
However, we would like to also stress the theoretical importance of having
a general pQCD framework that includes heavy quarks and 
is valid to all orders in perturbation theory
over a wide range of hard energy scales.

\begin{figure}[t]
\centerline{
\includegraphics[clip,width=0.65\textwidth]{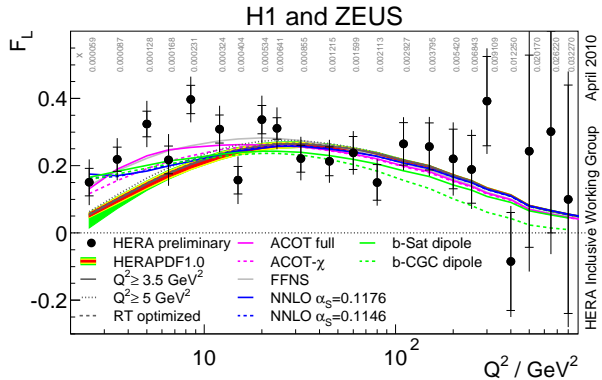}}
\caption{$F_{L}$ vs. $Q$ from combined HERA-I inclusive deep inelastic
cross sections measured by the H1 and ZEUS collaborations.
Figure taken from Ref.~\cite{hera:FL}.
\label{fig:slac-4-4}}
\end{figure}

An example, where higher order corrections are particularly important 
is the structure function $F_L$ in DIS.
The leading order (${\cal O}(\alpha_S^0)$) contribution to this structure function
vanishes for massless quarks due to helicity conservation (Callan-Gross relation).
This has several consequences:
%
1)~$F_L$ is useful for constraining the gluon PDF via the dominant subprocess $\gamma^* g \to q \bar q$.
2)~The heavy quark mass effects of order ${\cal O}(\tfrac{m^2}{Q^2})$ are relatively more 
pronounced.\footnote{%
Similar considerations also hold for target mass corrections (TMC) and higher twist terms.
We focus here mainly on the kinematic region $x<0.1$ where TMC are small \cite{Schienbein:2007gr}.
An inclusion of higher twist terms is beyond the scope of this study.}
%
3)~Since the first non-vanishing contribution to $F_L$ is  
next-to-leading order (up to mass effects), the 
NNLO and N$^3$LO corrections are more important than for $F_2$.
%
In Fig.~\ref{fig:slac-4-4} we show a comparison of different theoretical
calculations of $F_L$ with preliminary HERA data~\cite{hera:FL}.
As can be seen, in particular at small $Q^2$ (i.e.\ small $x$),
there are considerable differences between the predictions.

%

The purpose of this paper is to calculate the leading twist neutral current DIS structure functions
$F_2$ and $F_L$ in the ACOT factorization scheme up to order ${\cal O}(\alpha_S^3)$ (N$^3$LO)
and to estimate the error due to approximating the heavy quark mass terms 
${\cal O}(\alpha_S^2 \times \tfrac{m^2}{Q^2})$ 
and ${\cal O}(\alpha_S^3 \times \tfrac{m^2}{Q^2})$ 
in the higher order corrections.
The results of this study form the basis for using the ACOT scheme in NNLO global analyses
and for future comparisons with precision data for DIS structure functions.


This paper is organized as follows.
In Sec.~\ref{sec:schemes} we review theoretical approaches to include heavy flavors
in QCD calculations. Particular emphasis is put on the ACOT scheme which is a
minimal extension of the $\msbar$ scheme.
In Sec.~\ref{sec:ho} we present the  prescription for constructing
the approximate DIS structure functions in the ACOT scheme up to ${\cal O}(\alpha_S^3)$
order.
The corresponding numerical results are presented in Sec.~\ref{sec:numerics}.
Finally, in Sec.~\ref{sec:conclusion} we summarize the main results.
This work is based on  Ref.~\cite{Stavreva:2012bs}, and further details
can be found therein.

\section{Review of Theoretical Methods}
\label{sec:schemes}

We review theoretical methods which have been advanced to improve
existing QCD calculations of heavy quark production, and the impact
on recent experimental results.

\subsection{ACOT Scheme \label{subsec:acot}}

The ACOT renormalization scheme~\cite{Aivazis:1993kh,Aivazis:1993pi} provides a mechanism
to incorporate the heavy quark mass into the theoretical calculation
of heavy quark production both kinematically and dynamically. In 1998
Collins~\cite{Collins:1998rz} extended the factorization theorem to
address the case of heavy quarks; this work provided the theoretical
foundation that allows us to reliably compute heavy quark processes
throughout the full kinematic realm.

If we consider
the DIS production of heavy quarks at ${\cal O}(\alpha_{S}^{1})$
this involves the LO $QV\to Q$ process and the NLO $gV\to Q\bar{Q}$
process.%
\footnote{At NLO, there are corresponding quark-initiated terms; for simplicity
we do not display them here, but they are fully contained in our calculations
\protect\cite{Kretzer:1998ju}. %
}
The key ingredient provided by the ACOT scheme is the subtraction
term (SUB) which removes the {}``double counting'' arising from
the regions of phase space where the LO and NLO contributions overlap.
Specifically, at NLO order, we can express the total result as a sum
of
\begin{equation}
\sigma_{TOT}=\sigma_{LO}+\left\{ \sigma_{NLO}-\sigma_{SUB}\right\}
\label{eq:acot}
\end{equation}
where the subtraction term for the gluon-initiated processes is
\begin{equation}
\sigma_{SUB}=f_{g}\otimes\tilde{P}_{g\to Q}\otimes\sigma_{QV\to Q}.
\label{eq:sub}
\end{equation}
$\sigma_{SUB}$ represents a gluon emitted from a proton ($f_{g}$)
which undergoes a collinear splitting to a heavy quark $(\tilde{P}_{g\to Q})$
convoluted with the LO quark-boson scattering $\sigma_{QV\to Q}$.
Here, $\tilde{P}_{g\to Q}(x,\mu)=\frac{\alpha_{s}}{2\pi}\,\ln(\mu^{2}/m^{2})\, P_{g\to Q}(x)$
where $P_{g\to Q}(x)$ is the usual $\overline{MS}$ splitting kernel,
$m$ is the quark mass and $\mu$ is the renormalization scale
which we typically choose to be $\mu=Q$.

An important feature of the ACOT scheme is that it reduces to the
appropriate limit both as $m\to0$ and $m\to\infty$ as we illustrate
below.
%
Specifically, in the limit where the quark $Q$ is relatively heavy compared
to the characteristic energy scale $(\mu\lsim m)$, we find $\sigma_{LO}\sim\sigma_{SUB}$
such that $\sigma_{TOT}\sim\sigma_{NLO}$. In this limit, the ACOT
result naturally reduces to the Fixed-Flavor-Number-Scheme (FFNS)
result. In the FFNS, the heavy quark is treated as being extrinsic
to the hadron, and there is no corresponding heavy quark PDF ($f_{Q}\sim0$);
thus $\sigma_{LO}\sim0$. We also have $\sigma_{SUB}\sim0$ because
this is proportional to $\ln(\mu^{2}/m^{2})$. Thus, when the
quark $Q$ is heavy relative to the characteristic energy scale $\mu$,
the ACOT result reduces to $\sigma_{TOT}\sim\sigma_{NLO}$.

Conversely, in the limit where the quark $Q$ is relatively light compared
to the characteristic energy scale $(\mu\gsim m)$, we find that
$\sigma_{LO}$ yields the dominant part of the result, and the ``formal''
NLO ${\cal O}(\alpha_{S})$ contribution $\left\{ \sigma_{NLO}-\sigma_{SUB}\right\} $
is an ${\cal {\cal O}}(\alpha_{S})$ correction.
In this limit, the ACOT result will reduce to the
$\msbar$ Zero-Mass Variable-Flavor-Number-Scheme (ZM-VFNS) limit
exactly without any finite renormalizations. The quark
mass $m$ no longer plays any dynamical role and purely serves
as a regulator. The $\sigma_{NLO}$ term diverges due to the internal
exchange of the quark $Q$, and this singularity is canceled by
$\sigma_{SUB}$.

%

We illustrate the versatile role of the quark mass in
Fig.~\ref{fig:sacot} where we display $F_2^c$ as a function of $Q$
calculated in the  ZM-VFNS, FFNS, ACOT, and S-ACOT schemes.
We see that the ACOT scheme coincides with the FFNS for small $Q$,
and the  ZM-VFNS for large $Q$.
In Fig.~\ref{fig:nloLimit} we plot
$F_2^c$ as a function of the quark mass $m$ for a fixed $Q=10$~GeV
for the $\msbar$ ZM-VFNS and ACOT schemes.
We observe that when $m$ is within a decade or two of $\mu$, 
the quark mass plays a dynamic role; however, for $m\ll\mu$,
the quark mass purely serves as a regulator and the specific value
is not important. Operationally, it means we can obtain the $\msbar$
ZM-VFNS result either by i) computing the terms using dimensional
regularization and setting the regulator to zero, or ii) by computing
the terms using the quark mass as the regulator and then setting this
to zero.

The ACOT scheme is minimal in the sense that the construction of the
massive short distance cross sections does not need any
observable--dependent extra contributions or any regulators to smooth
the transition between the high and low scale regions.  The ACOT
prescription is: a) calculate the massive partonic cross sections,
and b) perform the factorization using the quark mass as regulator.

It is in this sense that we claim the ACOT scheme is the minimal
massive extension of the $\msbar$ ZM-VFNS. In the limit $m/\mu\to0$
it reduces exactly to the $\msbar$ ZM-VFNS, in the limit $m/\mu\gsim1$
the heavy quark decouples from the PDFs and we obtain exactly the FFNS
for $m/\mu\gg1$ and no finite renormalizations are needed.

\subsection{S-ACOT}

\begin{figure}
\begin{centering}
\subfloat[]{
\includegraphics[clip,width=0.45\textwidth]{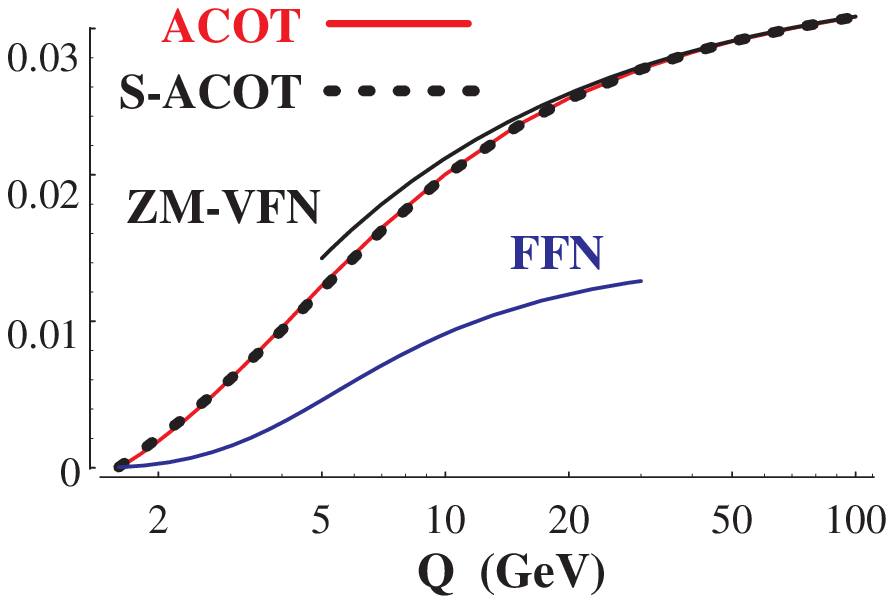}
\label{fig:sacot}
}
\subfloat[]{
\includegraphics[clip,width=0.45\textwidth]{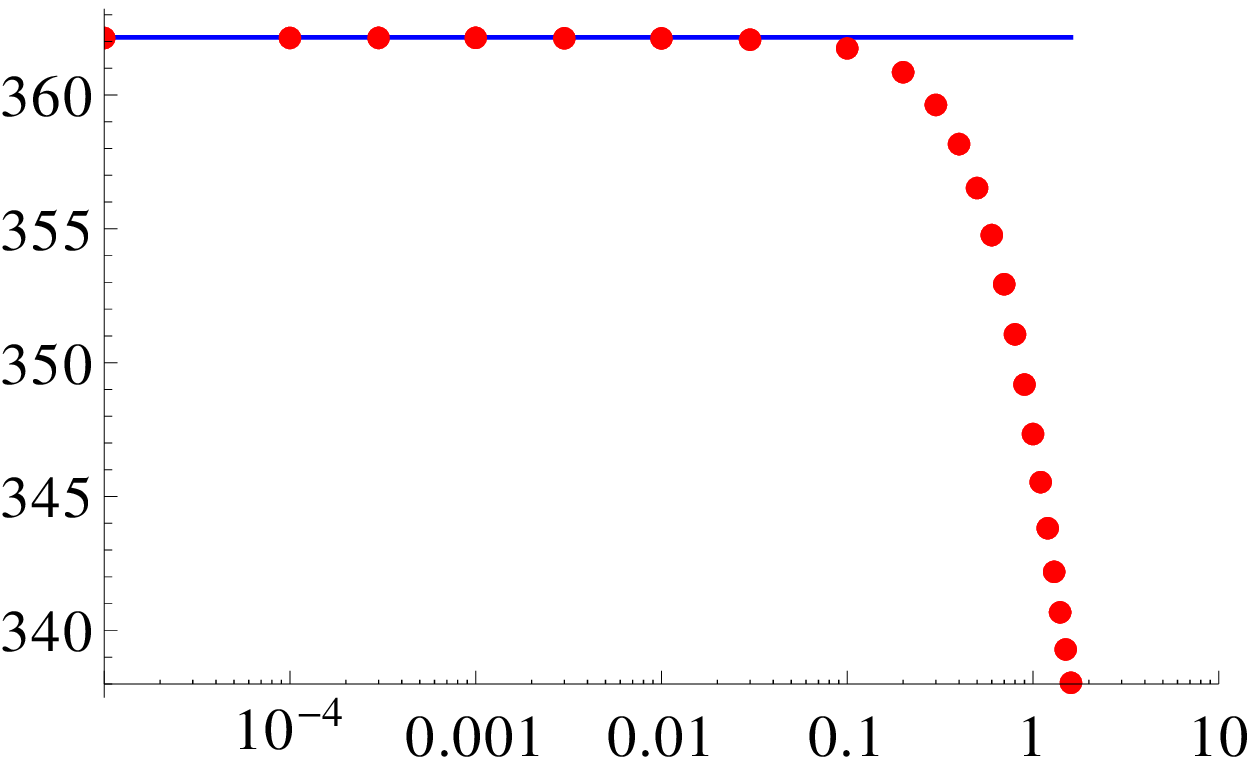}
\label{fig:nloLimit}
}
\caption{%
a)
$F_2^c$ for $x=0.1$ for NLO DIS heavy quark production
as a function of $Q$. We display calculations using the ACOT,
S-ACOT, Fixed-Flavor Number Scheme (FFNS),
and Zero-Mass Variable Flavor Number Scheme (ZM-VFNS).
The ACOT and S-ACOT results are virtually identical.
\protect\\
b) Comparison of $F_{2}^{c}(x,Q)$ (scaled by $10^4$) vs. 
the quark mass $m$ in GeV
for fixed $x=0.1$ and $Q=10$~GeV. The red dots are the full ACOT
result, and the blue line is the massless $\msbar$ result.
}
\label{fig:acotLimit}
\end{centering}
\end{figure}

In a corresponding application, it was observed that the heavy quark
mass could be set to zero in certain pieces of the hard scattering
terms without any loss of accuracy. This modification of the ACOT
scheme goes by the name Simplified-ACOT (S-ACOT) and can be summarized
as follows~\cite{Kramer:2000hn}.

\begin{quote}
\textbf{S-ACOT}: For hard-scattering processes with incoming
heavy quarks or with internal on-shell cuts on a heavy quark line,
the heavy quark mass can be set to zero ($m=0$) for these pieces.
\end{quote}

If we consider the case of NLO DIS heavy quark production, this means
we can set $m=0$ for the LO terms $\sigma_{QV\to Q}$ (incoming heavy quark),
and for the SUB terms (on-shell cut on an internal heavy quark line).
Hence, the only contribution which requires calculation with $m$ retained
is the NLO $gV\to Q\bar{Q}$ process. Figure~\ref{fig:sacot} displays
a comparison of a calculation using the ACOT scheme with all masses
retained vs. the S-ACOT scheme; as expected, these two results match
throughout the full kinematic region.

It is important to note that the S-ACOT scheme is not an approximation;
this is an exact renormalization scheme, extensible to all orders.

\subsection{ACOT and $\chi$-Rescaling}

As we have illustrated in Sec.~\ref{subsec:acot}, in the limit $Q^2\gg m^{2}$ the mass
simply plays the role of a regulator.
In contrast, for $Q^2 \sim m^{2}$ the value of the mass is of consequence for the physics.
The mass can enter dynamically in the hard-scattering matrix element,
and can enter kinematically in the phase space of the process.

We will demonstrate that for the processes of interest the primary
role of the mass is kinematic and not dynamic.
It was this idea which
was behind the original slow-rescaling prescription of \cite{Barnett:1976ak}
which considered DIS charm production (e.g., $\gamma c\to c)$ introducing
the shift
$
x\to \chi=x [1+(m_c/Q)^2].
$
This prescription accounted for the charm quark mass by effectively
reducing the phase space for the final state by an amount proportional
to $(m_{c}/Q)^{2}$.

This idea was extended in the $\chi$-scheme by realizing that (in
most cases) in addition to the observed final-state charm quark, there
is also an anti-charm quark in the beam fragments since all the charm
quarks are ultimately produced by gluon splitting ($g\to c\overline{c}$)
into a charm pair.
For this case the scaling variable becomes
$
\chi=x [1+(2m_c/Q)^2].
$
This rescaling is implemented in the ACOT$_{\chi}$ scheme, for
example~\cite{Amundson:1998zk,Tung:2001mv,Guzzi:2011ew}.%
\footnote{Use of more general rescaling prescriptions have been discussed
in Ref.~\cite{Nadolsky:2009ge}.
}
The factor $(1+(2m_{c})^{2}/Q^{2})$ represents a kinematic suppression
factor which will suppress the charm process relative to the lighter
quarks.
Additionally, the $\chi$-scaling ensures the threshold kinematics ($W^2>4m^2+M^2$) is satisfied; while it is important 
to satisfy this condition for large $x$, this may prove too restrictive  at small $x$ where 
the HERA data are especially precise.

To encompass all the above results, we can define a general scaling
variable $\chi(n)$ as
\begin{equation}
\chi(n)=x\left[1+\left(\frac{n\: m_{c}}{Q}\right)^{2}\right]
\label{eq:chin}
\end{equation}
where $n=\{0,1,2\}$. 
Here, $n=0$ corresponds to the massless result without rescaling, 
$n=1$ corresponds to the original Barnett slow-rescaling,
and $n=2$ corresponds to the $\chi$-rescaling.

\subsection{Phase Space (Kinematic) \& Dynamic Mass}
\label{sec:kinMass}

%
\begin{figure*}
\begin{centering}
\subfloat[
Comparison of $F_{2}^{c}(x,Q)$ vs. $Q$ for the NLO ACOT calculation for $x=\{10^{-1},10^{-3},10^{-5}\}$ (left to right) using zero dynamic mass {[}$\widehat{\sigma}(m=0)${]} to show the effect of $n$ scaling;
from top to bottom $n=\{0,1,2\}$ (pink, black, purple).]
{
\includegraphics[width=0.32\textwidth]{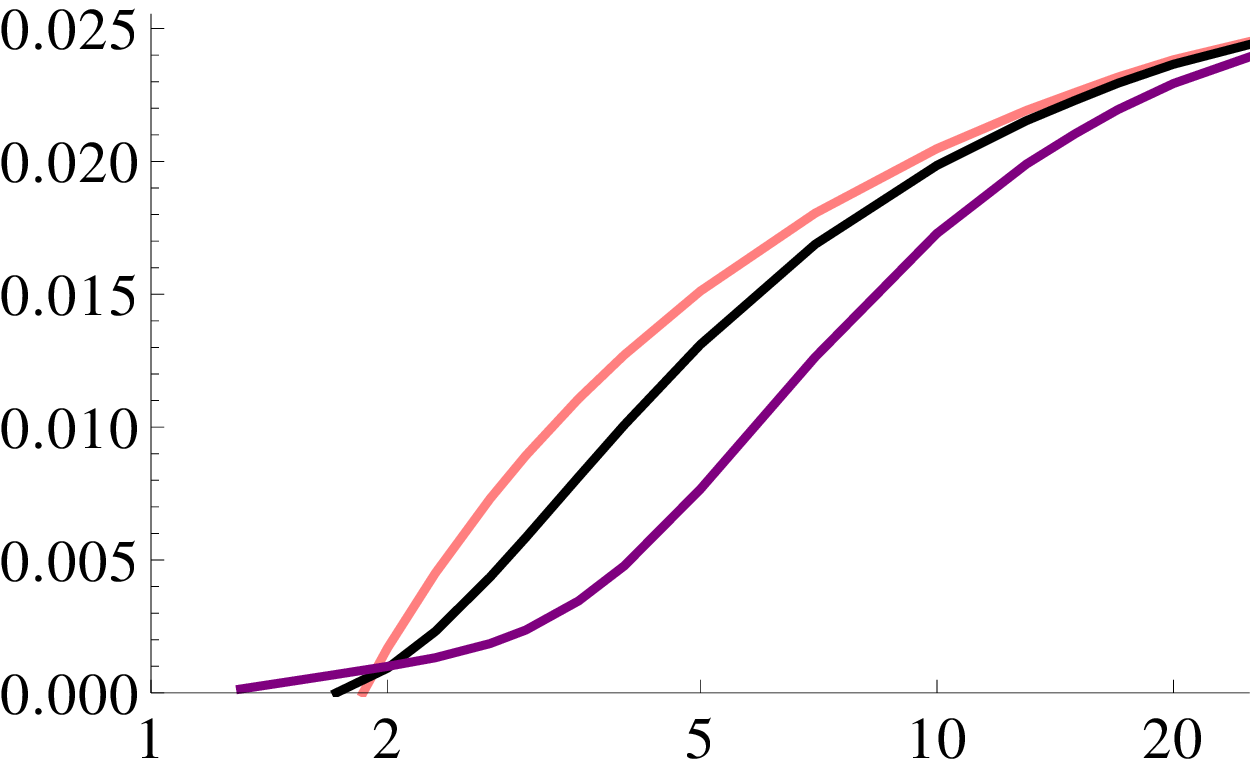}
\hfil
\includegraphics[width=0.32\textwidth]{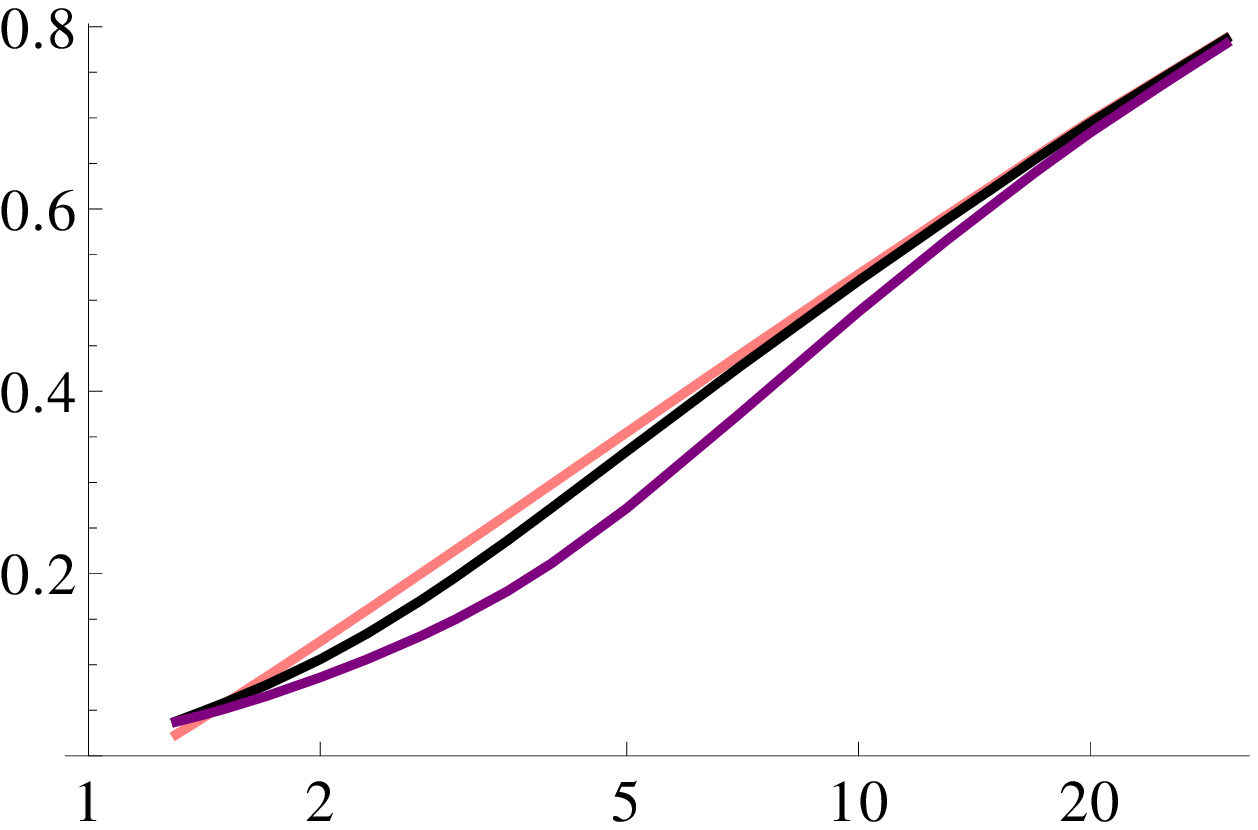}
\hfil
\includegraphics[width=0.32\textwidth]{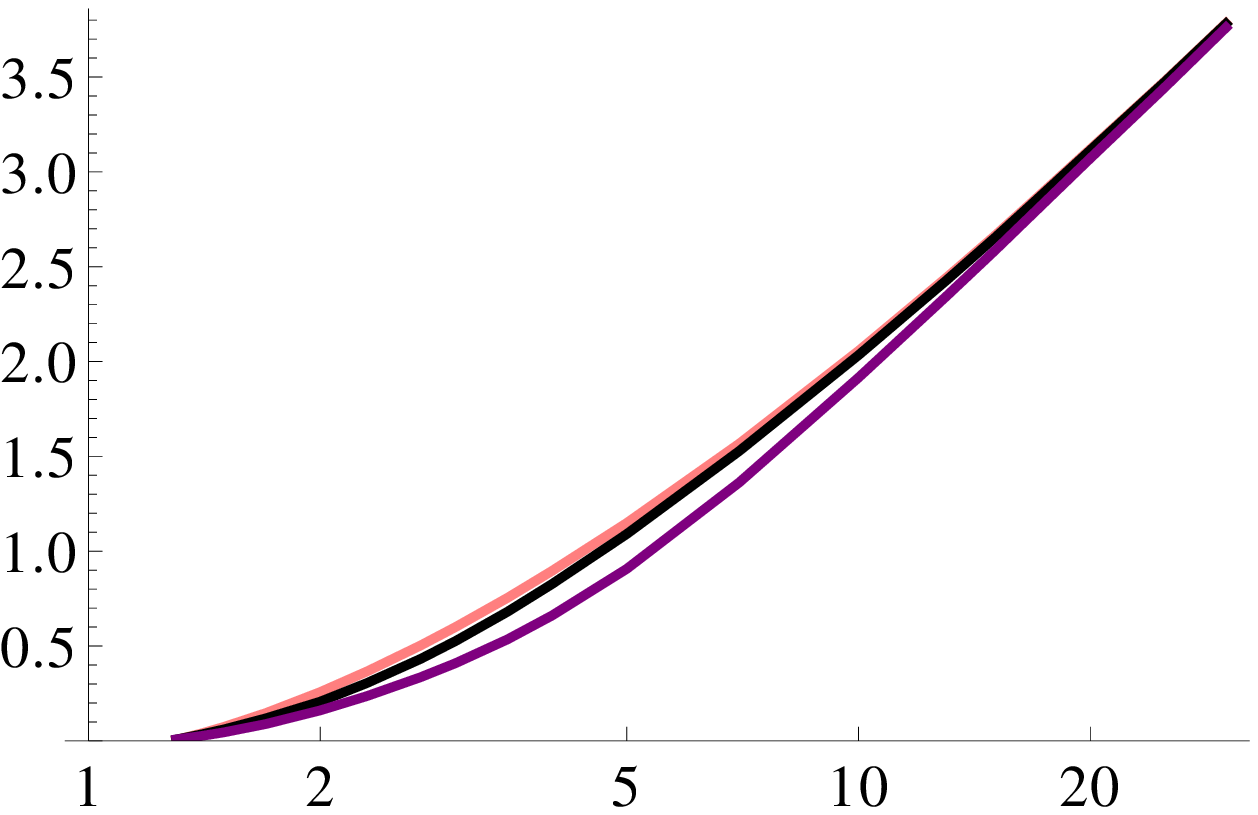}
\label{fig:nscale}
}

\subfloat[
Comparison of $F_{2}^{c}(x,Q)$ vs. $Q$ for the NLO ACOT calculation for
$x=\{10^{-1},10^{-3},10^{-5}\}$ (left to right).
Here we keep the scaling fixed $n=2$ and compare the effect of varying
the dynamic mass in the Wilson coefficient.
The upper (cyan) curve uses a non-zero dynamic mass {[}$\widehat{\sigma}(m=1.3)${]}
and the lower (purple) curve uses a zero dynamic mass {[}$\widehat{\sigma}(m=0)${]}.
]
{
\includegraphics[width=0.32\textwidth]{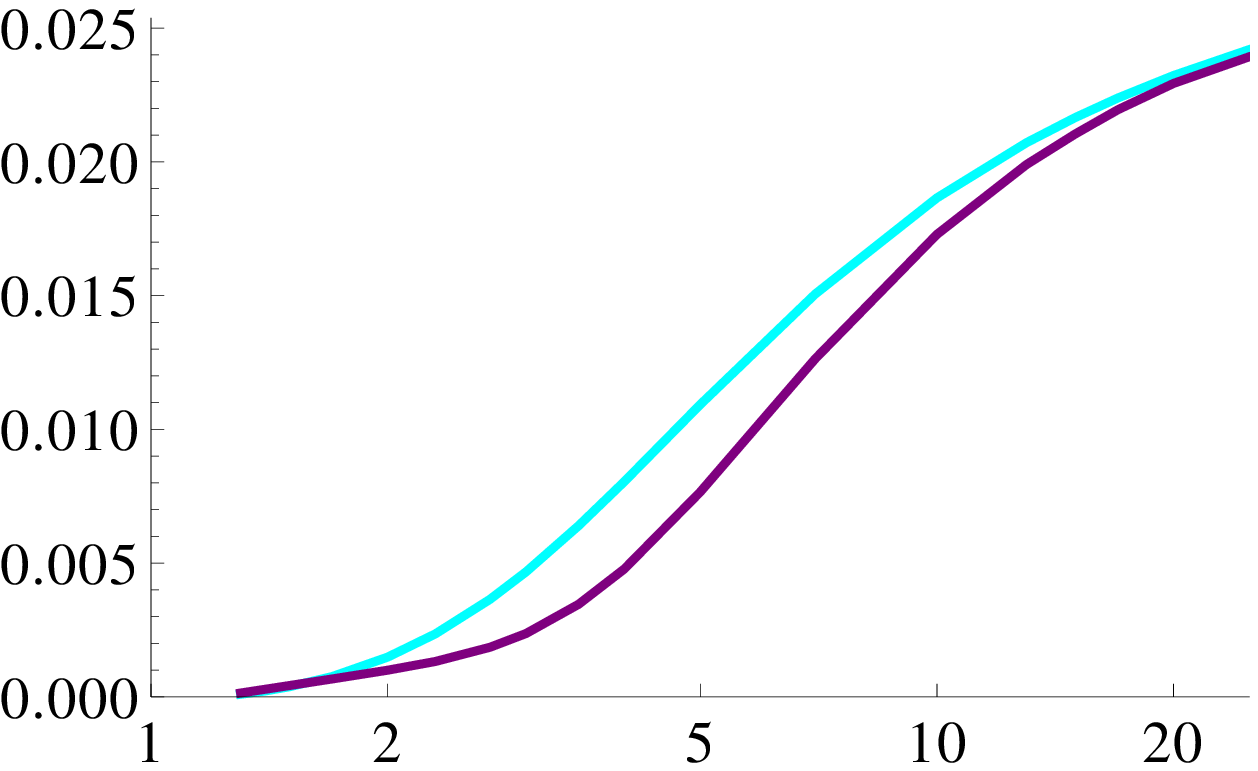}
\hfil
\includegraphics[width=0.32\textwidth]{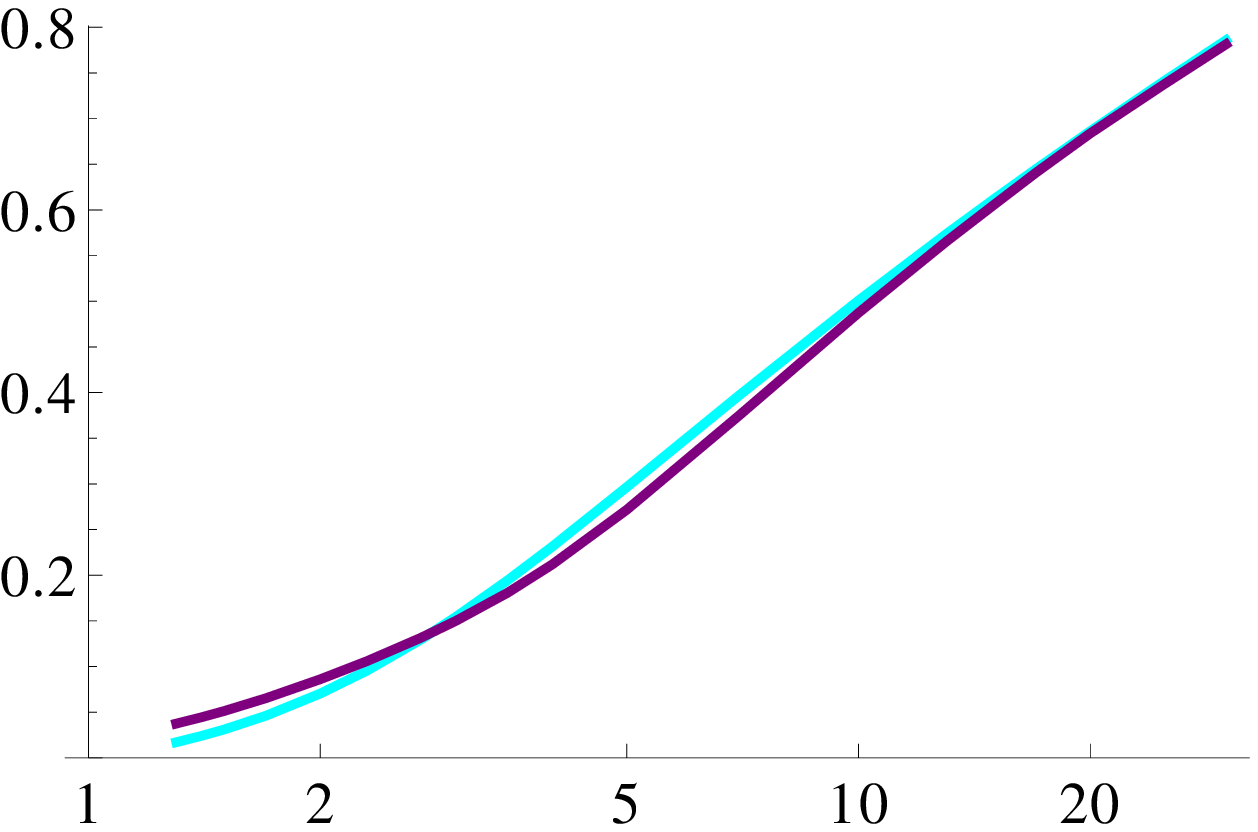}
\hfil
\includegraphics[width=0.32\textwidth]{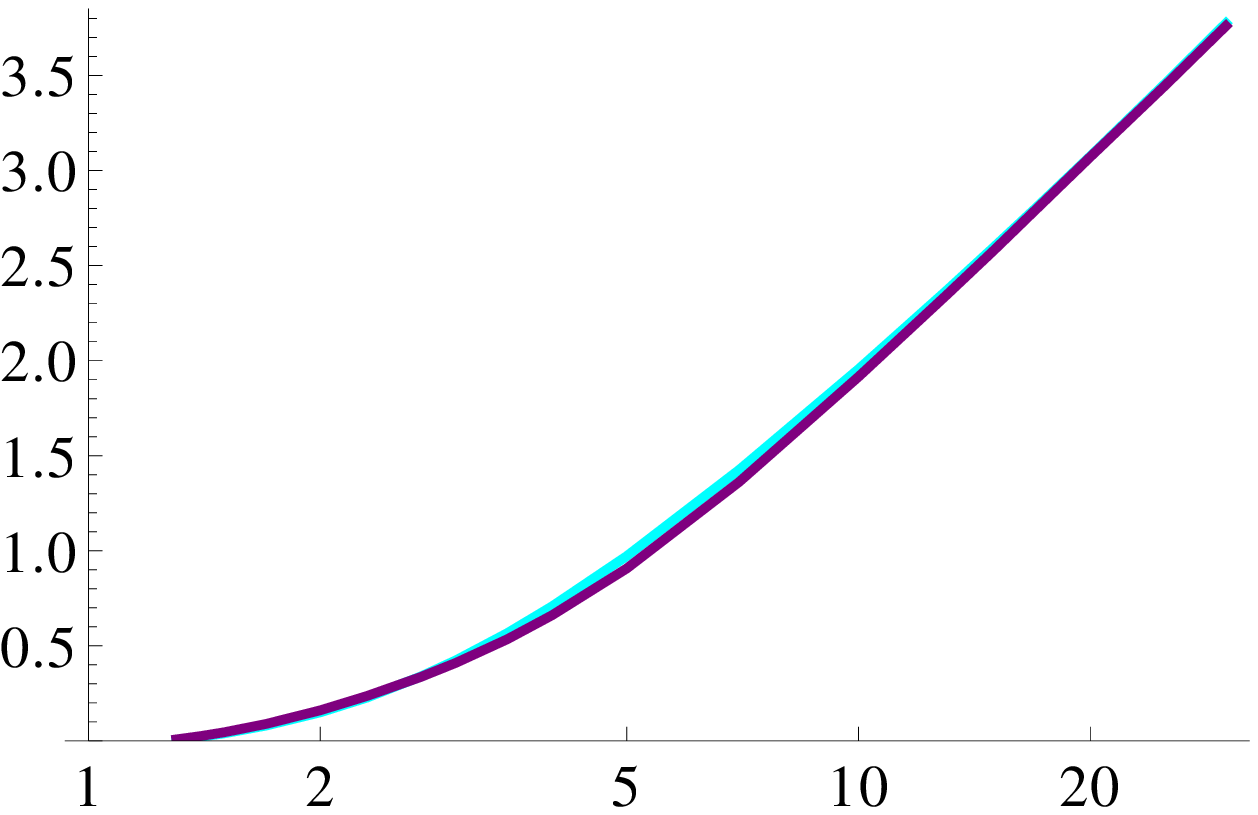}
\label{fig:dynamic}
}
\caption{Comparison of Phase Space (Kinematic) \& Dynamic Mass Effects}
\label{fig:massEffects}
\end{centering}
\end{figure*}

We now investigate the effects of separately varying the mass entering
the $\chi(n)$ variable taking into account the phase space constraints and
the mass value entering the hard scattering cross section $\widehat{\sigma}(m)$.
We call the former mass parameter ``phase space (kinematic) mass'' and the
latter ``dynamic mass''.

In Fig.~\ref{fig:nscale} we display $F_{2}^{c}(x,Q)$ vs. $Q$. 
The family of 3 curves shows the   NLO ACOT calculation
with $\chi(n)$ scaling using a zero dynamic mass for the hard scattering. 
We compare this with 
 Fig.~\ref{fig:dynamic} which shows
$F_2^c(x,Q)$ in the NLO ACOT scheme using a fixed  $n=2$ scaling, but 
varying the mass used in the  hard-scattering cross section.
The upper (cyan) curves use a non-zero dynamic mass {[}$\widehat{\sigma}(m_c=1.3)${]}
and the lower (purple) curves have been obtained with a vanishing dynamic mass {[}$\widehat{\sigma}(m_c=0)${]}.
We observe that the effect of the 'dynamic mass' in $\widehat{\sigma}(m_c)$ is only
of consequence in the limited region $Q\gsim m,$ and even in this
region the effect is minimal.
In contrast, the influence of the phase space
(kinematic) mass 
shown in  Fig.~\ref{fig:nscale}
is larger than the dynamic mass
shown in  Fig.~\ref{fig:dynamic}.

In conclusion, we have shown that (up to ${\cal O}(\alpha_S)$)
the phase space mass dependence is generally the dominant
contribution to the DIS structure functions.
Assuming that this observation remains true at higher orders, 
it is possible to obtain a good approximation of the structure functions
in the  ACOT scheme at NNLO and N$^3$LO using the massless Wilson 
coefficients together with a non-zero phase space mass entering via the
$\chi(n)$-prescription.

\subsection{Other massive schemes}

There are a number of other schemes for incorporating the
heavy quark mass terms, and we briefly note a few examples.
The Thorne-Roberts (TR) scheme~\cite{Thorne:1997ga,Thorne:1997uu} and its
derivatives (TR') are designed to provide a smooth threshold behavior, 
and this is implemented by including pieces of the higher order 
contributions. 
The FONLL scheme~\cite{Cacciari:1998it} was originally developed to
match fixed order calculations with resumed ones 
in the case of
heavy quark hadroproduction;
this approach has been generalized and applied to other applications including 
DIS structure functions \cite{Forte:2010ta}.
Details and comparisons of these approached is outlined in the 
2009 Les Houches Workshop report \cite{Binoth:2010ra}.

\section{ACOT scheme beyond NLO}
\label{sec:ho}

In Sec.~\ref{sec:kinMass} we have shown using the  NLO  full ACOT scheme
that the dominant mass effects 
are those coming from the phase space which can be  taken into account
via a generalized slow-rescaling $\chi(n)$-prescription.
Assuming that a similar relation remains true at higher orders one can construct
the following approximation to the full ACOT result  up to N$^3$LO ($\Ord(\alpha_S^3)$):
\begin{equation}
{\rm ACOT} [\Ord(\alpha_S^{0+1+2+3})]
\simeq
{\rm ACOT} [\Ord(\alpha_S^{0+1})] + \ZMVFNS_\chi [\Ord(\alpha_S^{2+3})].
\end{equation}
Here, the massless Wilson coefficients at ${\cal O}(\alpha\,\alpha_{S}^{2})$
and ${\cal O}(\alpha\,\alpha_{S}^{3})$ are substituted for the 
 Wilson coefficients in the ACOT  scheme as the corresponding massive coefficients 
have not yet been computed.

There has been a calculation of neutral current electroproduction 
(equal quark masses, vector coupling) 
of 
heavy quarks at this order 
by Smith \& VanNeerven \protect\cite{Laenen:1992zk} 
in the FFNS which could be used to obtain the massive Wilson coefficients in the 
S-ACOT scheme by applying appropriate collinear subtraction terms;%
\footnote{For the 
original ACOT scheme it would then still be necessary to compute the
massive Wilson coefficients for the heavy quark initiated subprocess at 
${\cal O}(\alpha\,\alpha_{S}^{2})$.
See Refs.~\cite{Guzzi:2011ew,Guzzi:2011dk} for details.
}
however, this is beyond the scope of this paper.
For charge current case massive calculations are available
at order ${\cal O}(\alpha\,\alpha_{S})$~\cite{PhysRevD.23.56,Gluck:1996ve,Blumlein:2011zu}
and partial results at order
${\cal O}(\alpha\,\alpha_{S}^2)$~\cite{Buza:1997mg}.

Here, we argue  that the massless Wilson coefficients at
${\cal O}(\alpha\,\alpha_{S}^{2})$ together with a $\chi(n)$-prescription
provide a very good approximation of the exact result.
At worst, the maximum error would be of order $\Ord(\alpha\, \alpha_{S}^{2}\times[m^{2}/Q^{2}])$.
However, based on the arguments of Sec.~\ref{sec:kinMass} we expect the inclusion
of the phase space mass effects to contain the dominant higher order contributions
so that the actual error should be substantially smaller.

The massless higher order coefficient functions for the DIS structure
function $F_2$ via photon exchange can be found in
Refs.~\cite{Furmanski:1981cw,Bardeen:1978yd,Altarelli:1978id,vanNeerven:1991nn,Zijlstra:1991qc,Zijlstra:1992qd,vanNeerven:1999ca,vanNeerven:2000uj,Vermaseren:2005qc,Moch:2002sn}.
The expressions for the structure function $F_L$ have been calculated
in Refs.~\cite{SanchezGuillen:1990iq,Zijlstra:1991qc,vanNeerven:1999ca,Moch:2004xu,Vermaseren:2005qc}.


We now consider our choice for the appropriate  generalized $\chi(n)$-rescaling variable.
For the purposes of this study, we will vary the phase space mass 
using the $\chi(n)$ rescaling with $n=\{0,1,2\}$.
While $n=0$ corresponds to the massless case (no rescaling), 
it is not obvious whether $n=1$ or $n=2$ is the preferred rescaling choice
for higher orders. Thus, we will use the range between  $n=1$ and  $n=2$
as a measure of our theoretical uncertainty arising from this ambiguity.

\section{Results}
\label{sec:numerics}


We now present our results for the $F_2$ and $F_L$ structure
functions calculated at N${}^3$LO in the extended ACOT scheme.
The initial PDFs, based on the Les Houches benchmark
set~\cite{Giele:2002hx} are evolved using the QCDNUM
program~\cite{Botje:2010ay}.
In the calculation we set $m_c=1.3$~GeV,  $m_b=4.5$~GeV and
$\alpha_s(M_Z)=0.118$.


\begin{figure*}
\begin{centering}
\subfloat[$F_{2}$ vs. $Q$.
\label{fig:f2x135}]{
\includegraphics[width=0.32\textwidth]{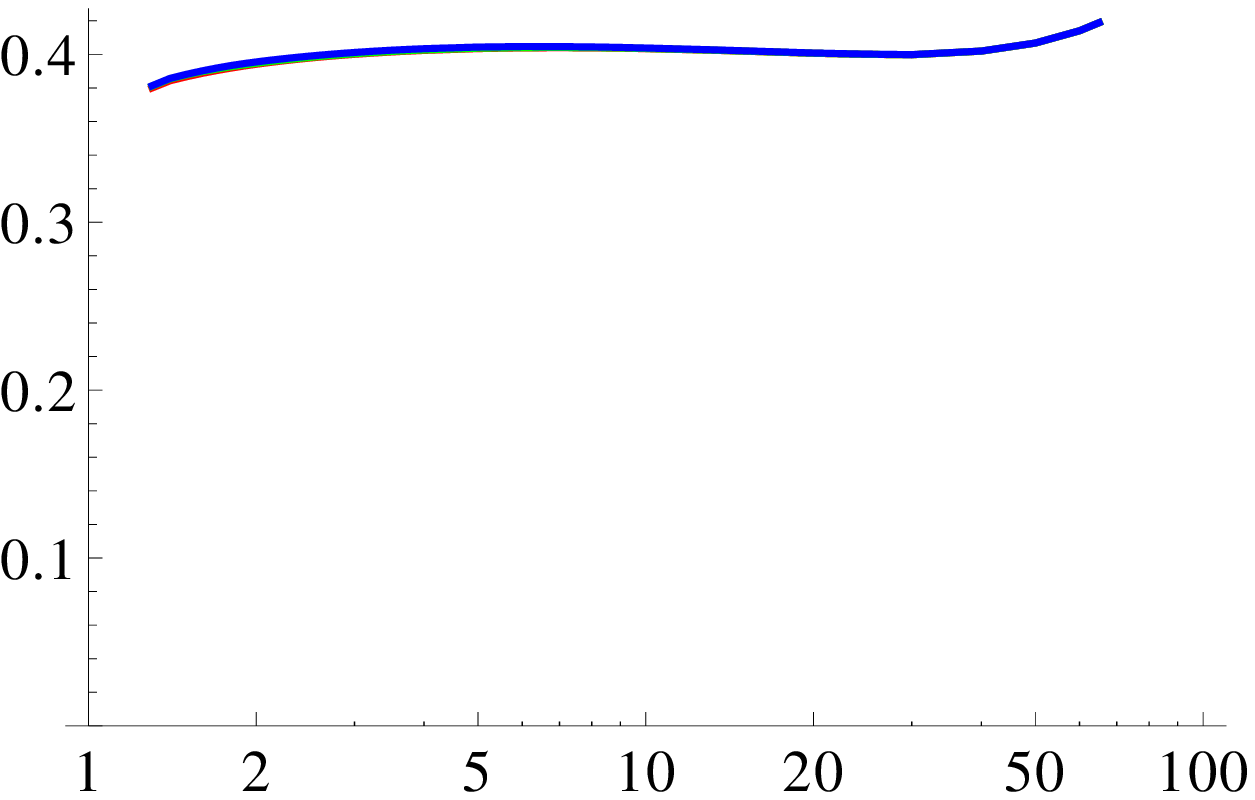}
\hfil
\includegraphics[width=0.32\textwidth]{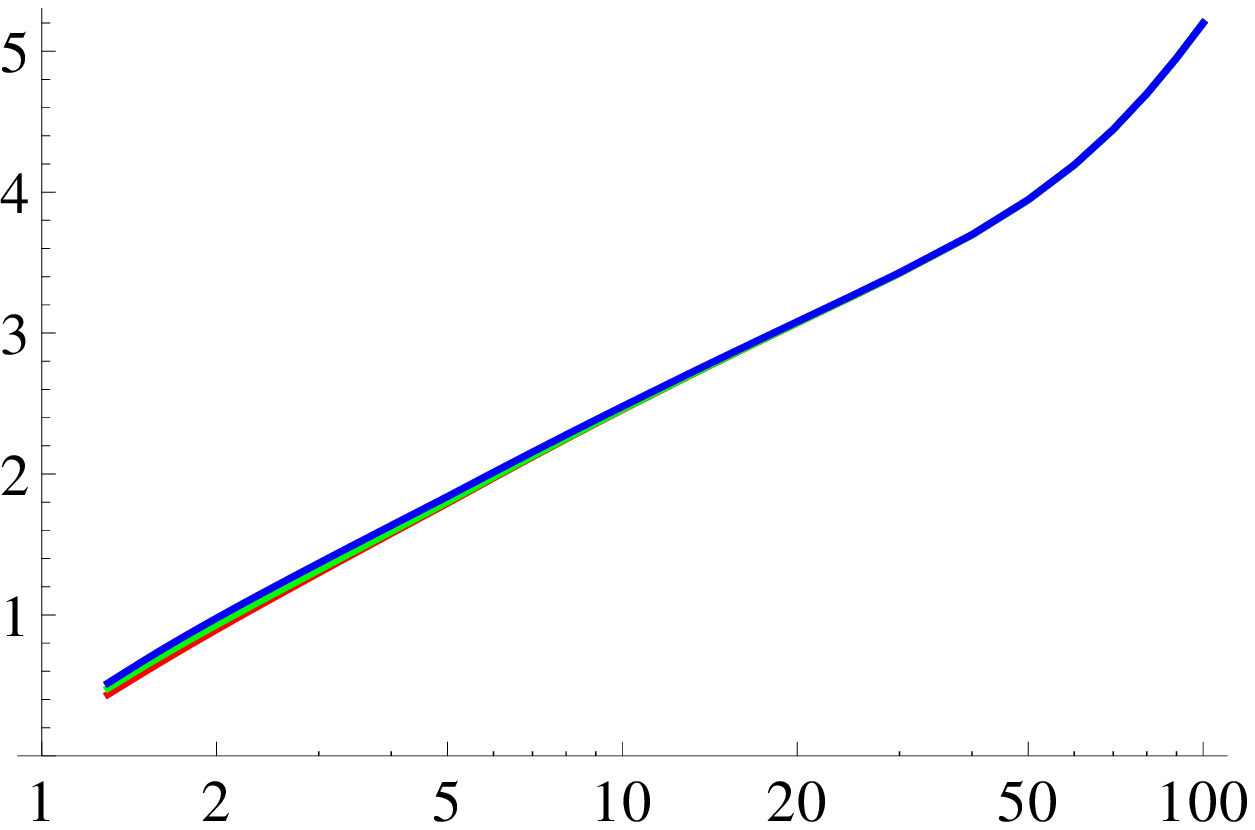}
\hfil
\includegraphics[width=0.32\textwidth]{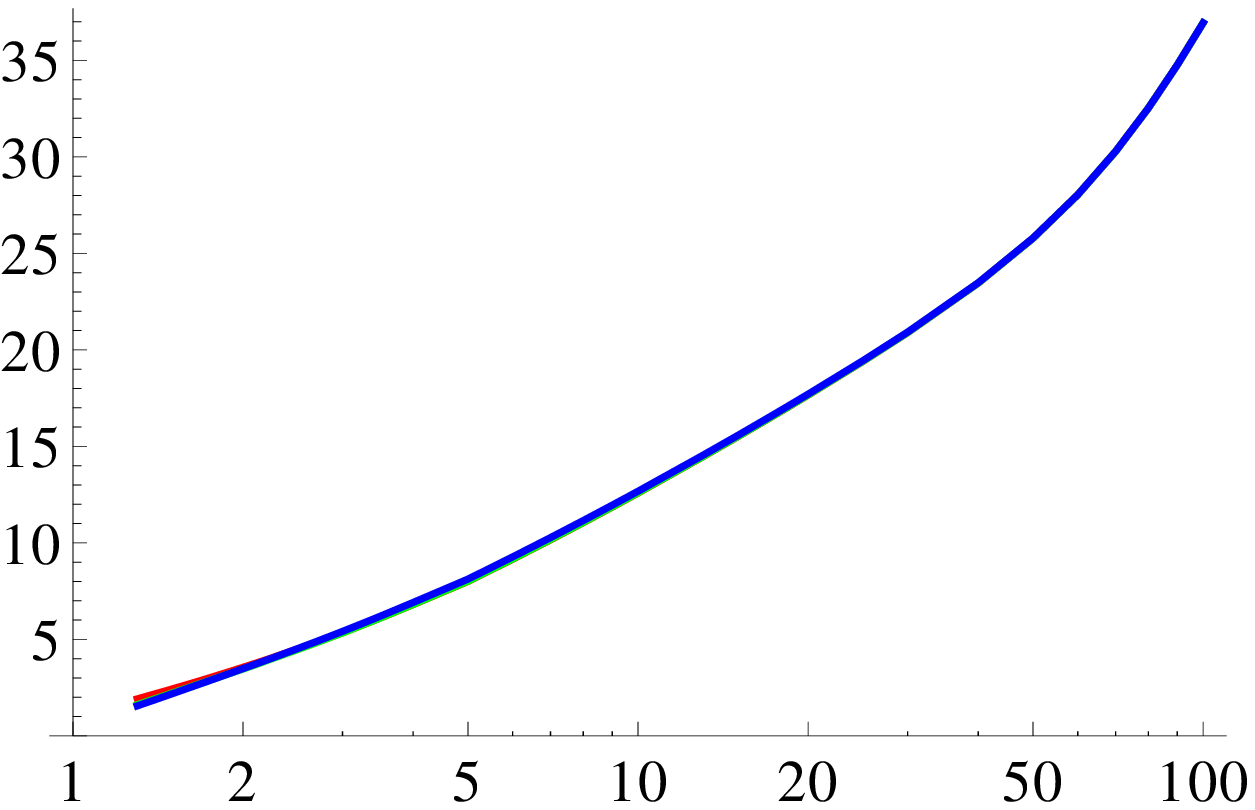}
}

\subfloat[$F_{L}$ vs. $Q$.
\label{fig:fLx135}]{
\includegraphics[width=0.32\textwidth]{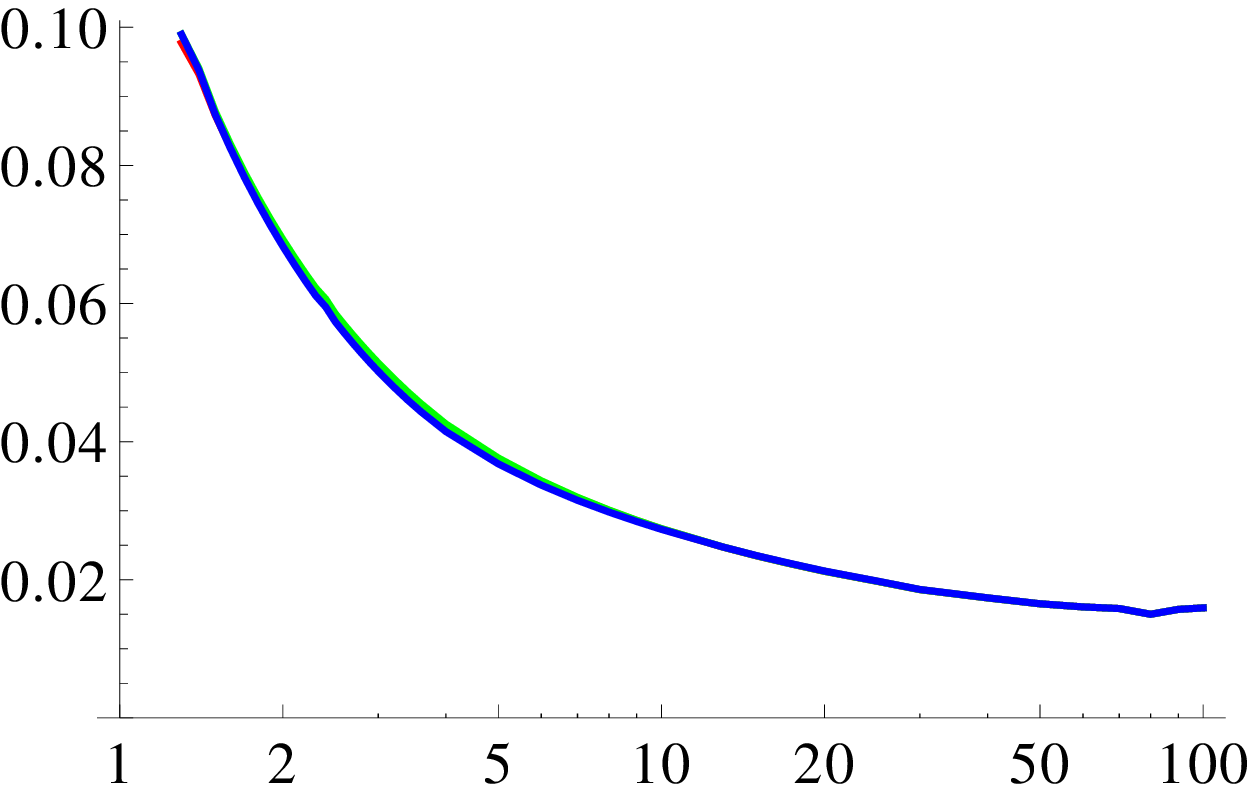}
\hfil
\includegraphics[width=0.32\textwidth]{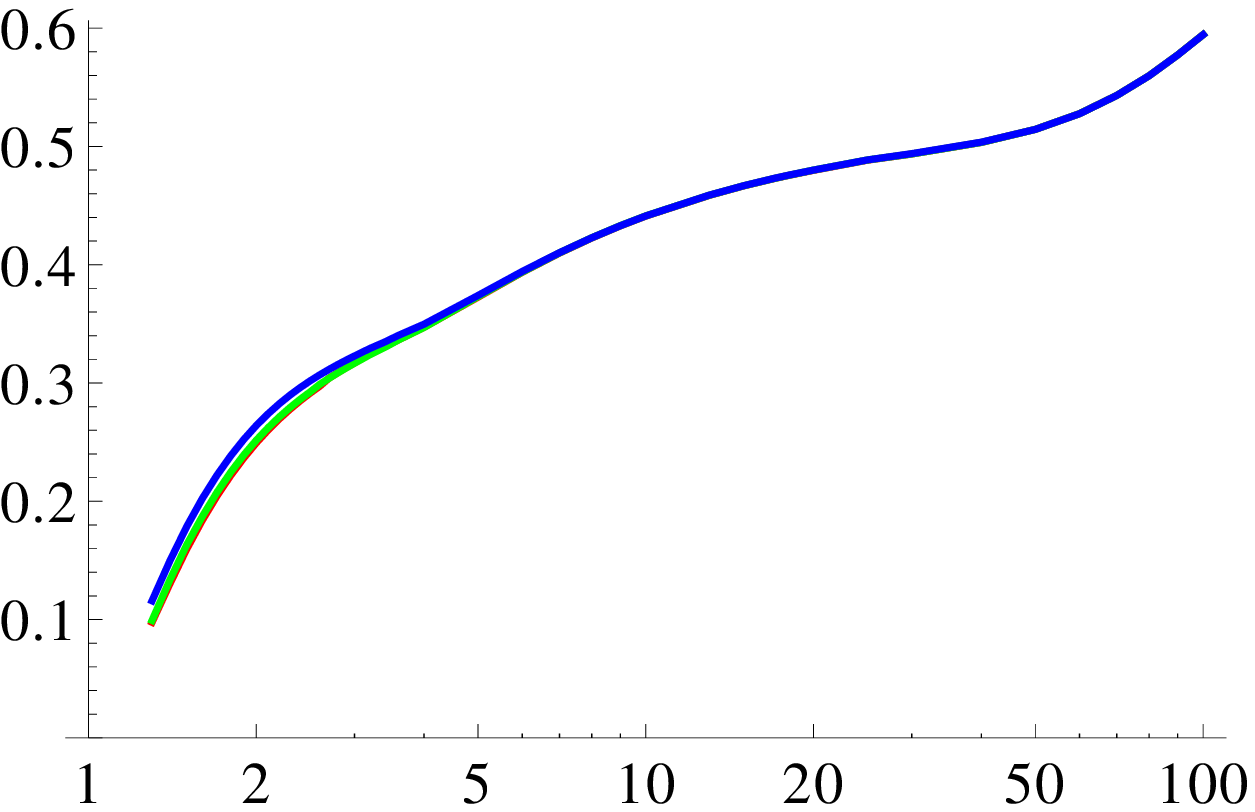}
\hfil
\includegraphics[width=0.32\textwidth]{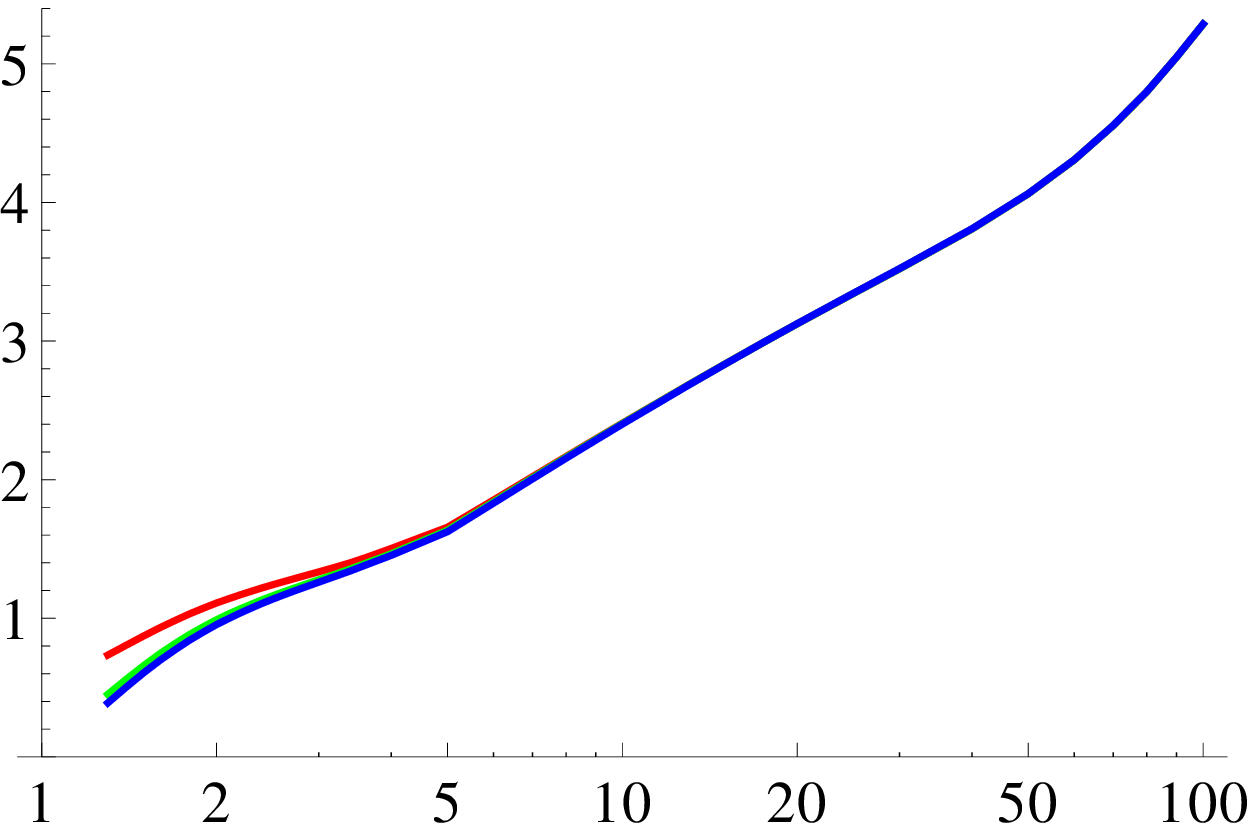}
}
\caption{$F_{2,L}$ vs. $Q$ at N$^3$LO for fixed $x=\{10^{-1},10^{-3},10^{-5}\}$
(left to right). The three lines show the scaling variable: $n=\{0,1,2\}$
(red, green, blue). We observe the effect of the $n$-scaling is negligible
except for very small $Q$ values.}
\end{centering}
\end{figure*}

%

In Figures~\ref{fig:f2x135} and~\ref{fig:fLx135} we display the
structure functions $F_{2}$ and $F_{L}$, respectively, for selected
$x$ values as a function of $Q$. Each plot has three curves which
are computed using $n$-scalings of $\{0,1,2\}$. We observe that
the effect of the $n$-scaling is negligible except for very small $Q$
values. This result is in part because the heavy quarks are only a fraction
of the total structure function, and the effects of the $n$-scaling
are reduced at larger $Q$ values.

In Ref.~\cite{Stavreva:2012bs} we magnify the small $Q$ region of $F_L$ of
Fig.~\ref{fig:fLx135} for $x=10^{-5}$, where the effects of using different
scalings are largest. We can see that for  inclusive observables, the  $n=1$ and $n=2$
scalings give nearly identical results, but they differ from the massless
case ($n=0$). 
This result,  together with the observation that at NLO kinematic mass
effects are dominant, suggests that the error we have in our approach
is relatively small and approximated  by the band between $n=1$ and $n=2$ results.


\begin{figure*}
\begin{centering}
\subfloat[$F_2^j/F_2$ vs. $Q$.
\label{fig:f2RatN123}]{
\includegraphics[width=0.32\textwidth]{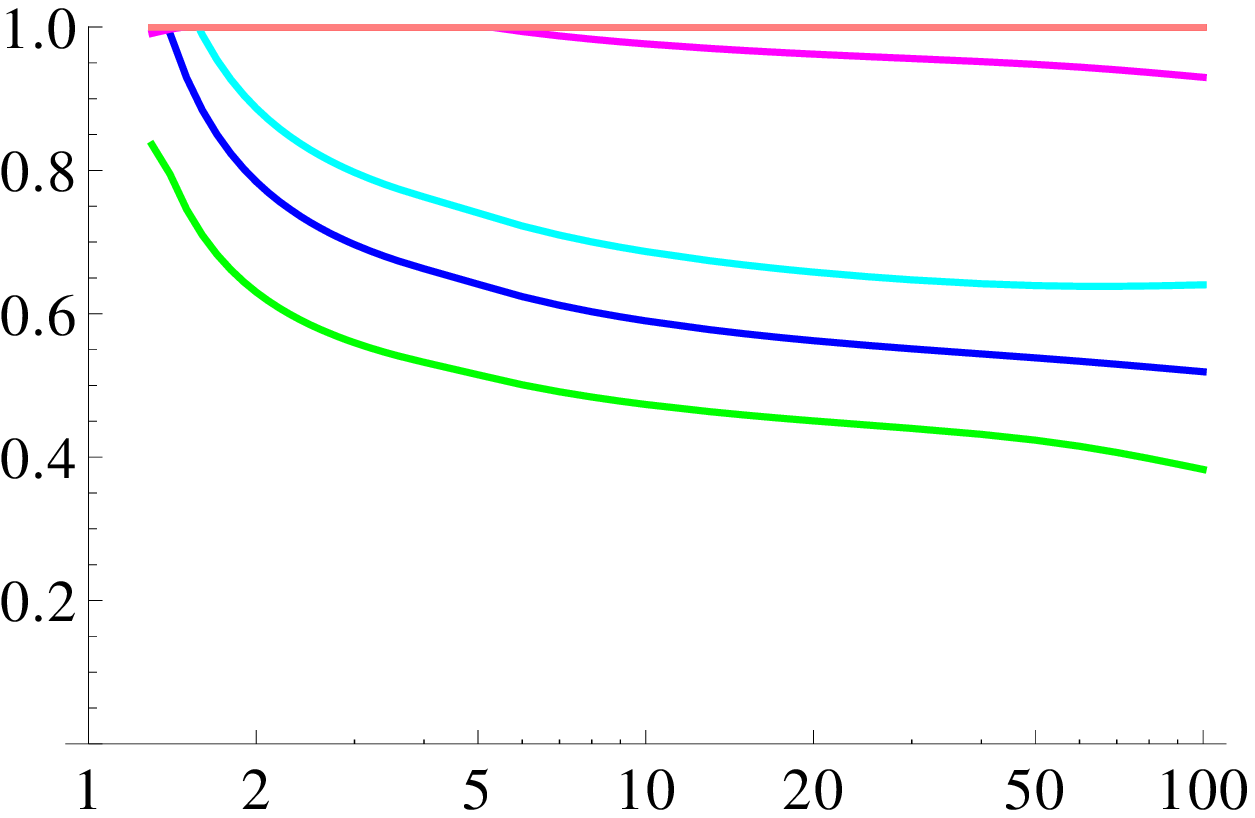}
\hfil
\includegraphics[width=0.32\textwidth]{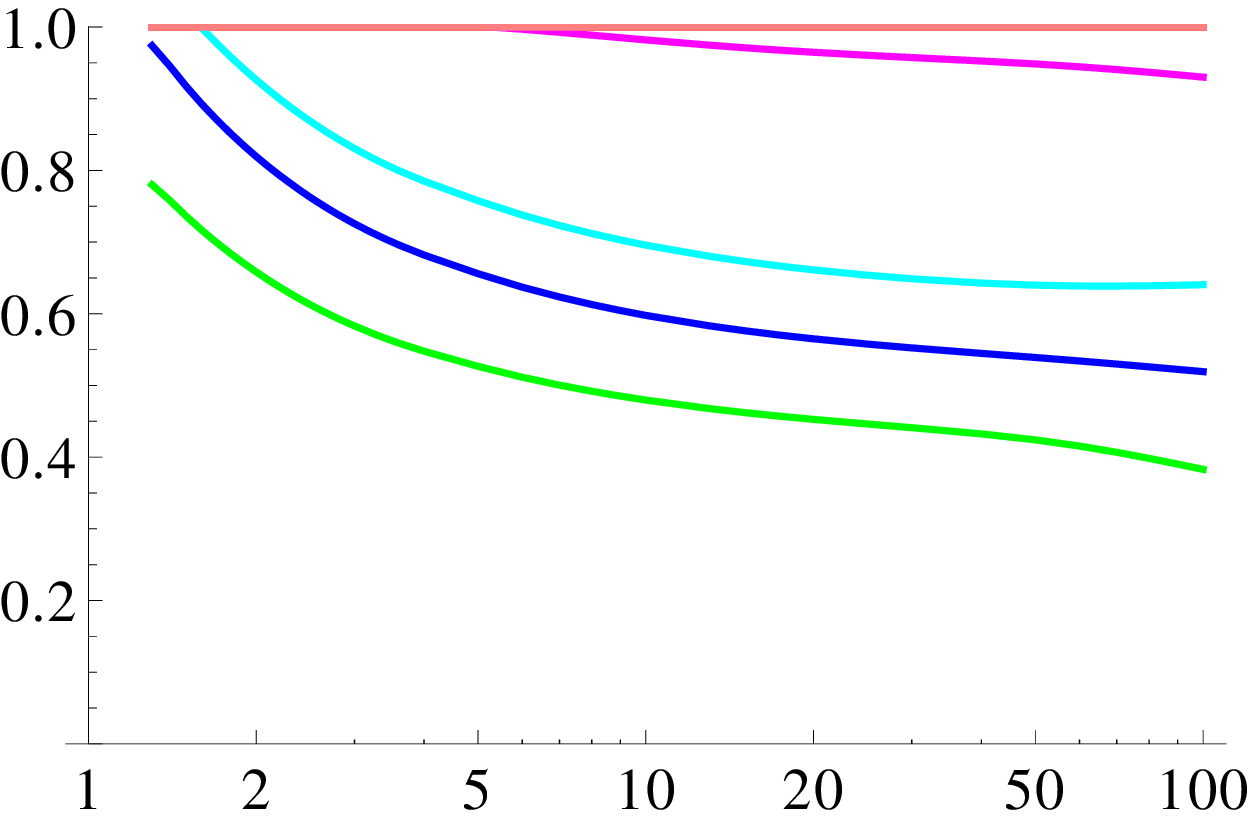}
\hfil
\includegraphics[width=0.32\textwidth]{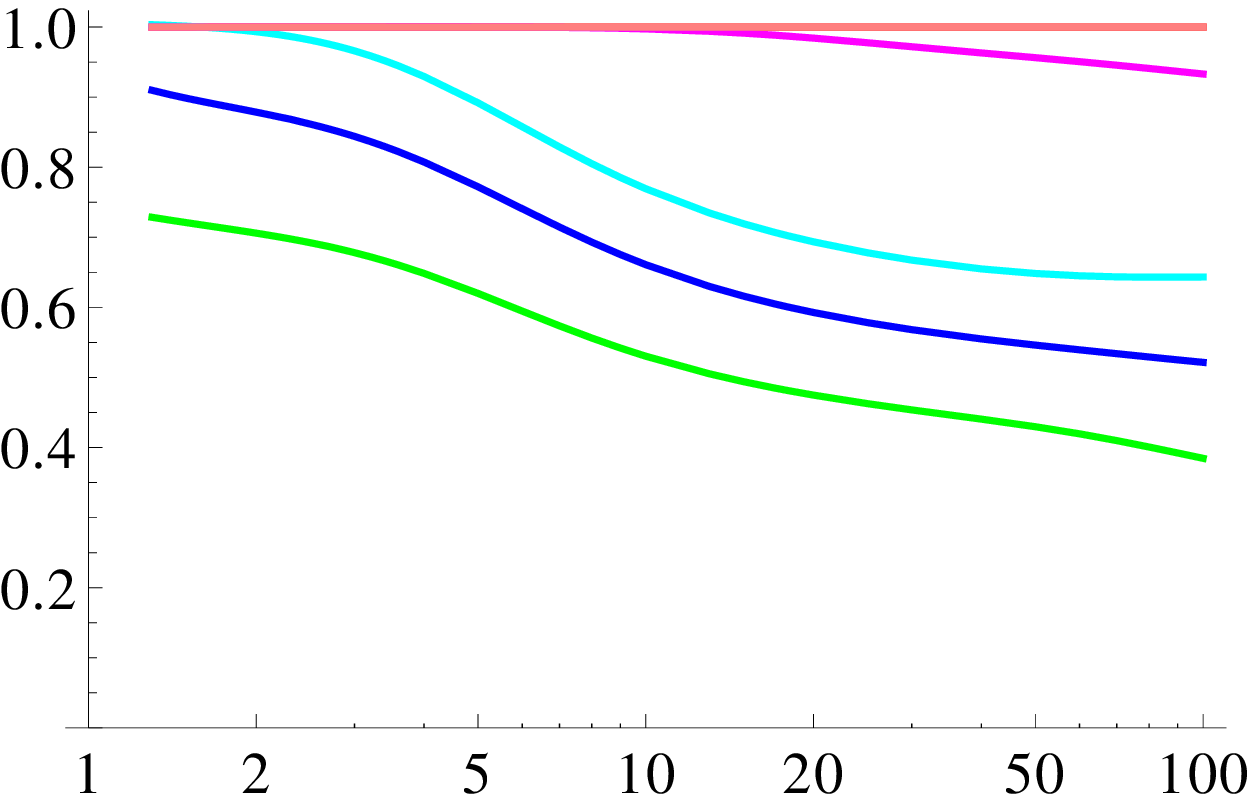}
}

\subfloat[$F_L^j/F_L$ vs. $Q$.
\label{fig:fLRatN123}]{
\includegraphics[width=0.32\textwidth]{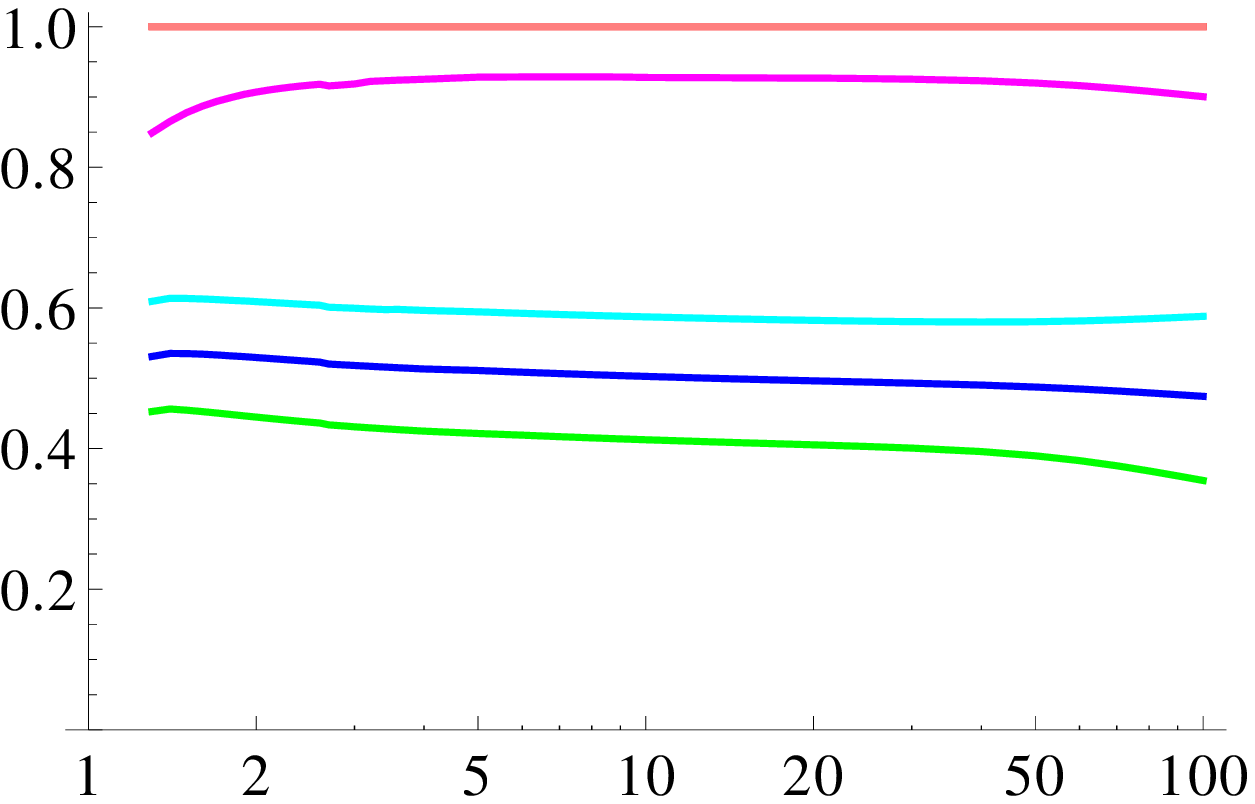}
\hfil
\includegraphics[width=0.32\textwidth]{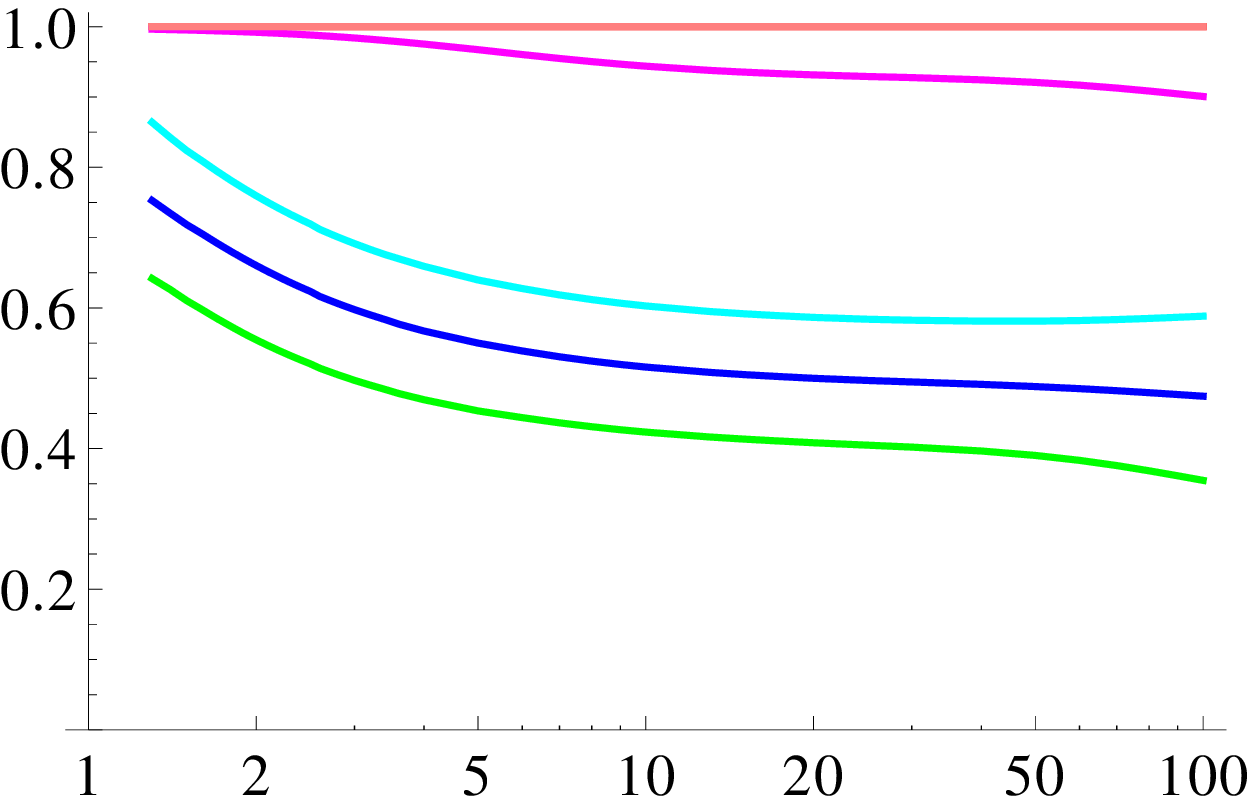}
\hfil
\includegraphics[width=0.32\textwidth]{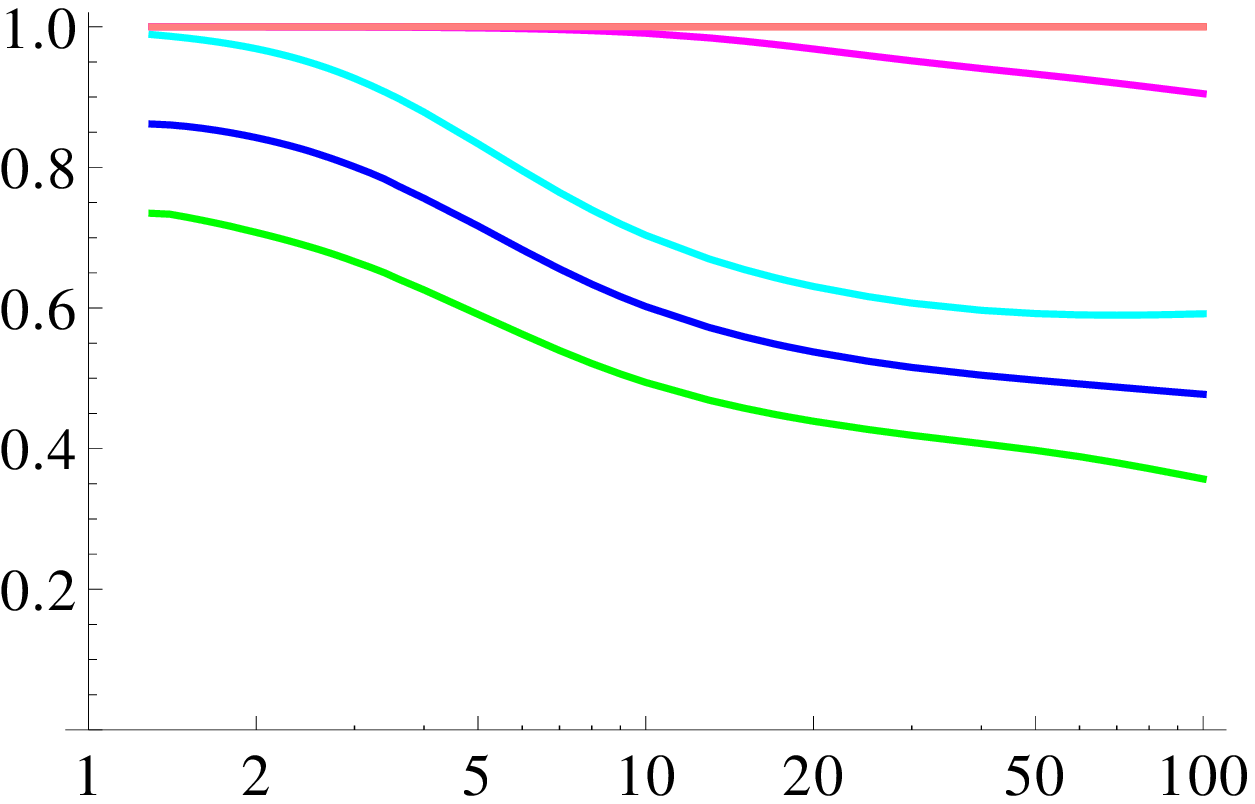}
}
\caption{
Effect of $\chi(n)$-scaling for $n=\{0,1,2\}$ (left to right)
at N$^3$LO for fixed $x=\{10^{-3}\}$.
Reading from the bottom we have fractional contribution for each
(final-state) quark flavor to $F_{2,L}^j/F_{2,L}$ vs. $Q$
from $\{u,d,s,c,b\}$ (green, blue, cyan, magenta, pink).}
\end{centering}
\end{figure*}

We can investigate the effects of the $\chi(n)$-scaling in more details by 
examining the flavor decomposition of the structure functions. 
In Figures~\ref{fig:f2RatN123} and~\ref{fig:fLRatN123} we display
the fractional contributions of quark flavors
to the structure functions $F_{2,L}$ for selected $n$-scaling values as a function
of $Q$. 
We observe the $n$-scaling reduces the relative contributions
of charm and bottom at low $Q$ scales. For example, without any $n$-scaling
($n=0$) we find the charm and bottom quarks contribute an unusually
large fraction  at very low scales $(Q\sim m_{c})$
as they are (incorrectly) treated as massless partons in this region.
The result of the different $n$-scalings ($n=1,2$) is to introduce
a kinematic penalty which properly suppresses the contribution of
these heavy quarks in the low $Q$ region. 
In the following, we will generally use the $n=2$ scaling for our comparisons.


\begin{figure*}
\begin{centering}
\subfloat[
$F_2^i/F_2$ vs. $Q$.
\label{fig:F2trans}
]
{
\includegraphics[width=0.32\textwidth]{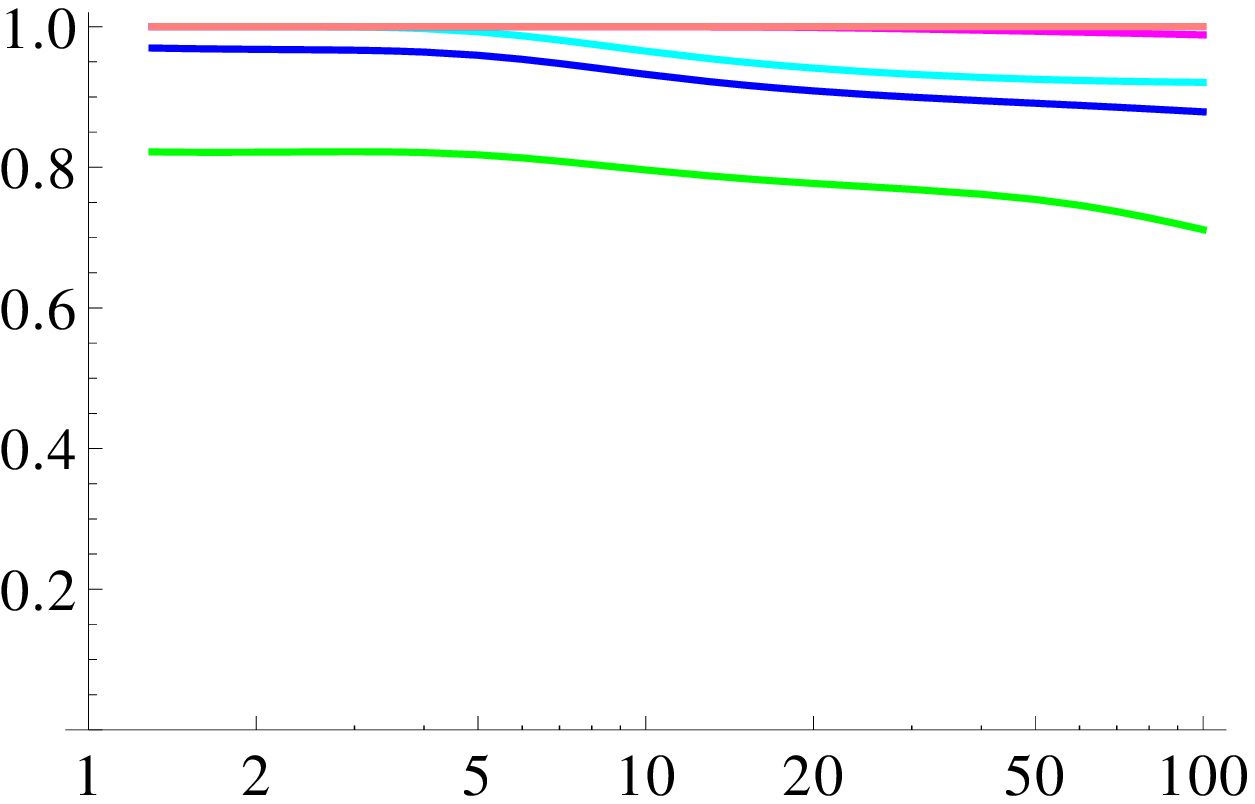}
\hfil
\includegraphics[width=0.32\textwidth]{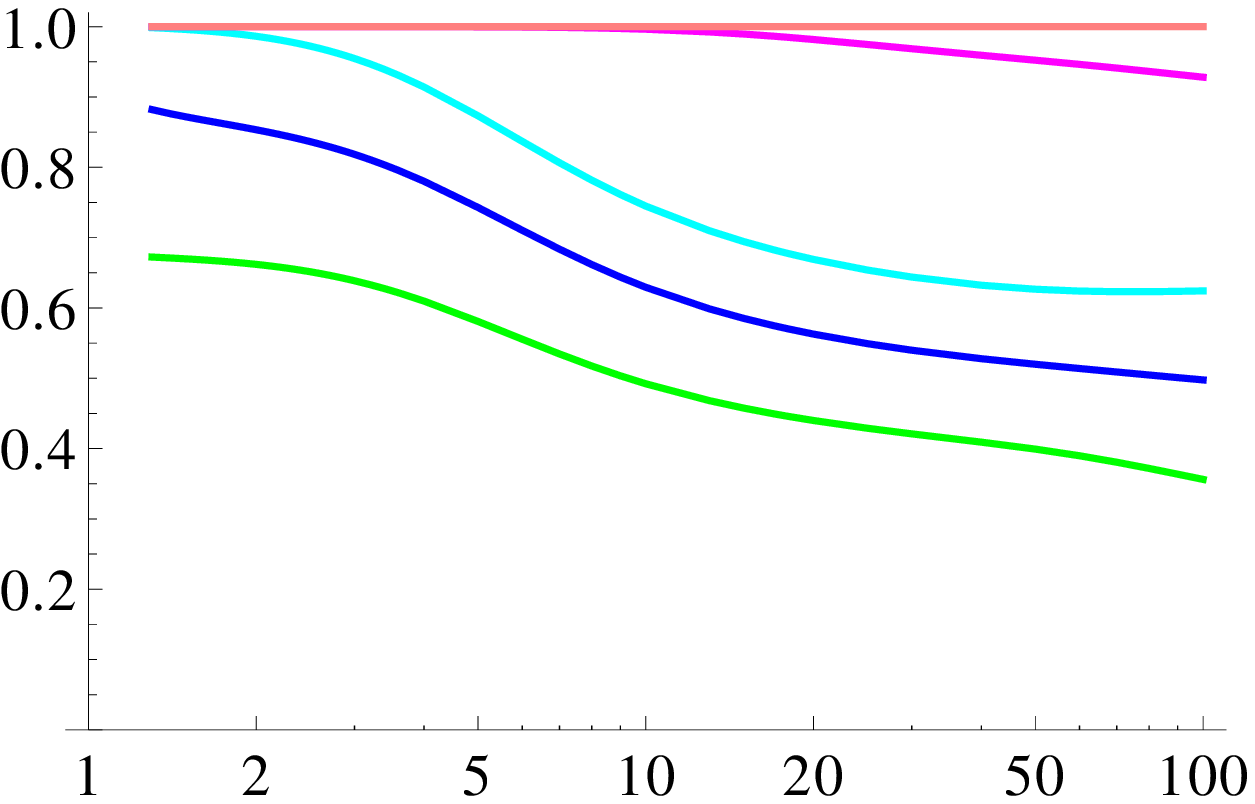}
\hfil
\includegraphics[width=0.32\textwidth]{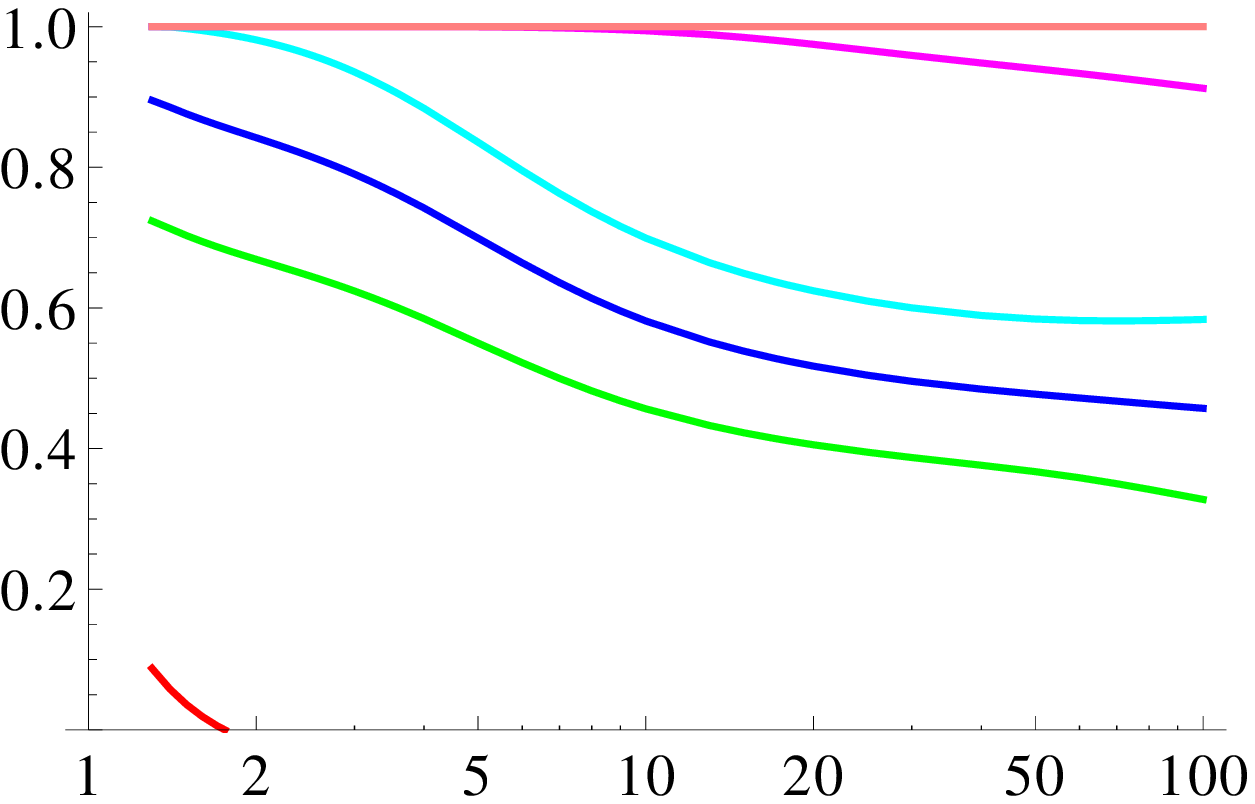}
}

\subfloat[
$F_L^i/F_L$ vs. $Q$.
\label{fig:FLtrans}
]
{
\includegraphics[width=0.32\textwidth]{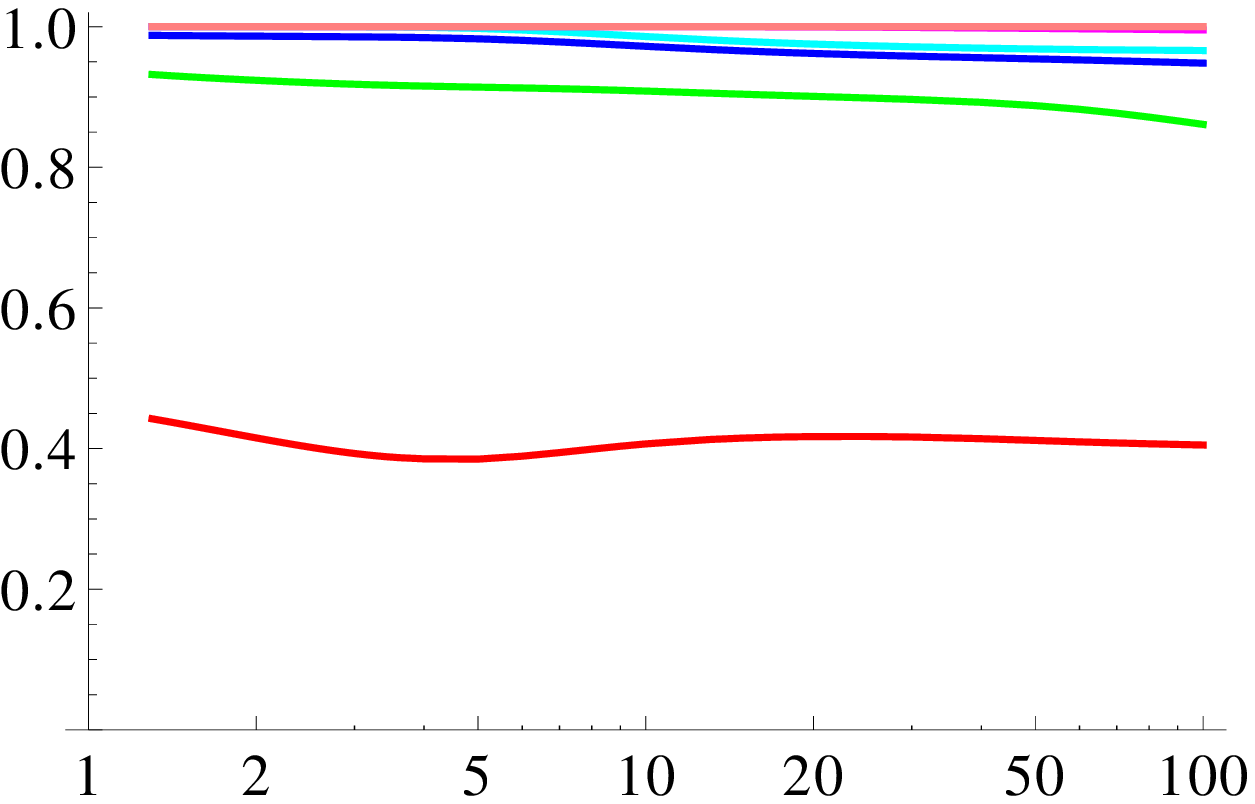}
\hfil
\includegraphics[width=0.32\textwidth]{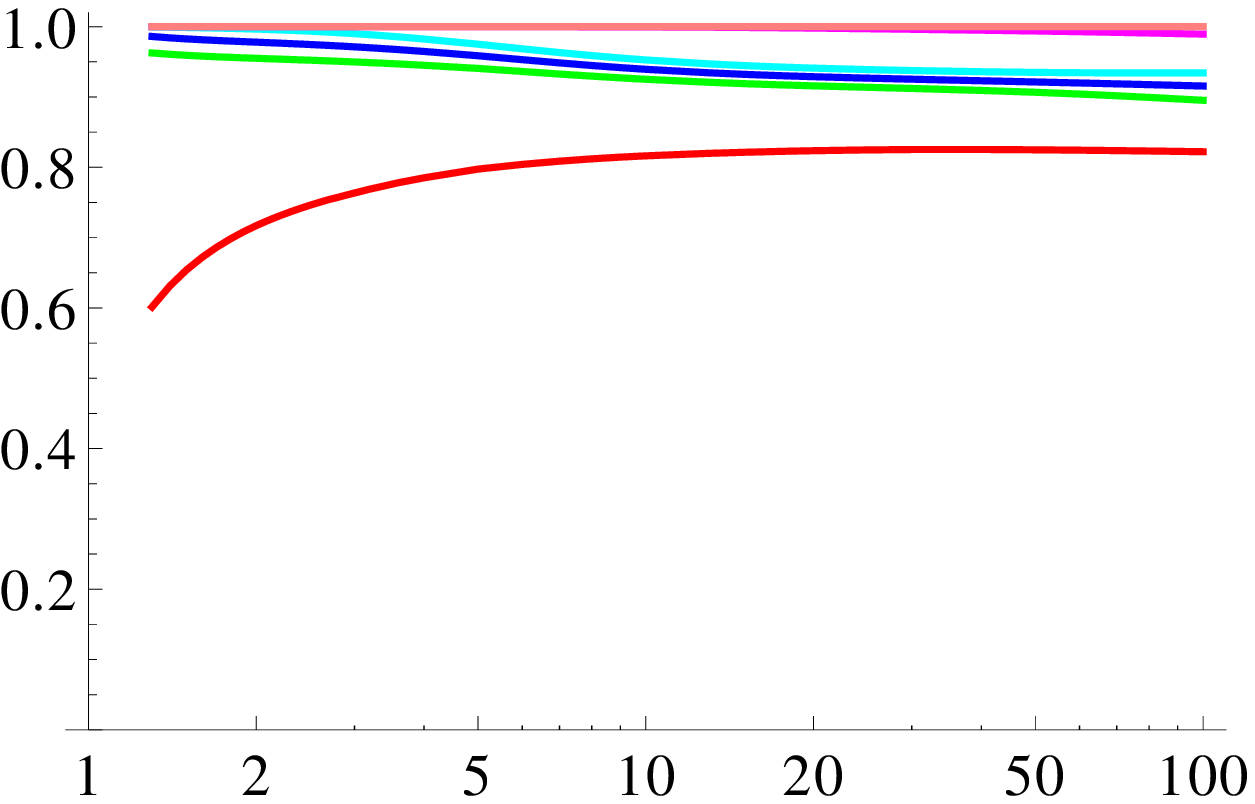}
\hfil
\includegraphics[width=0.32\textwidth]{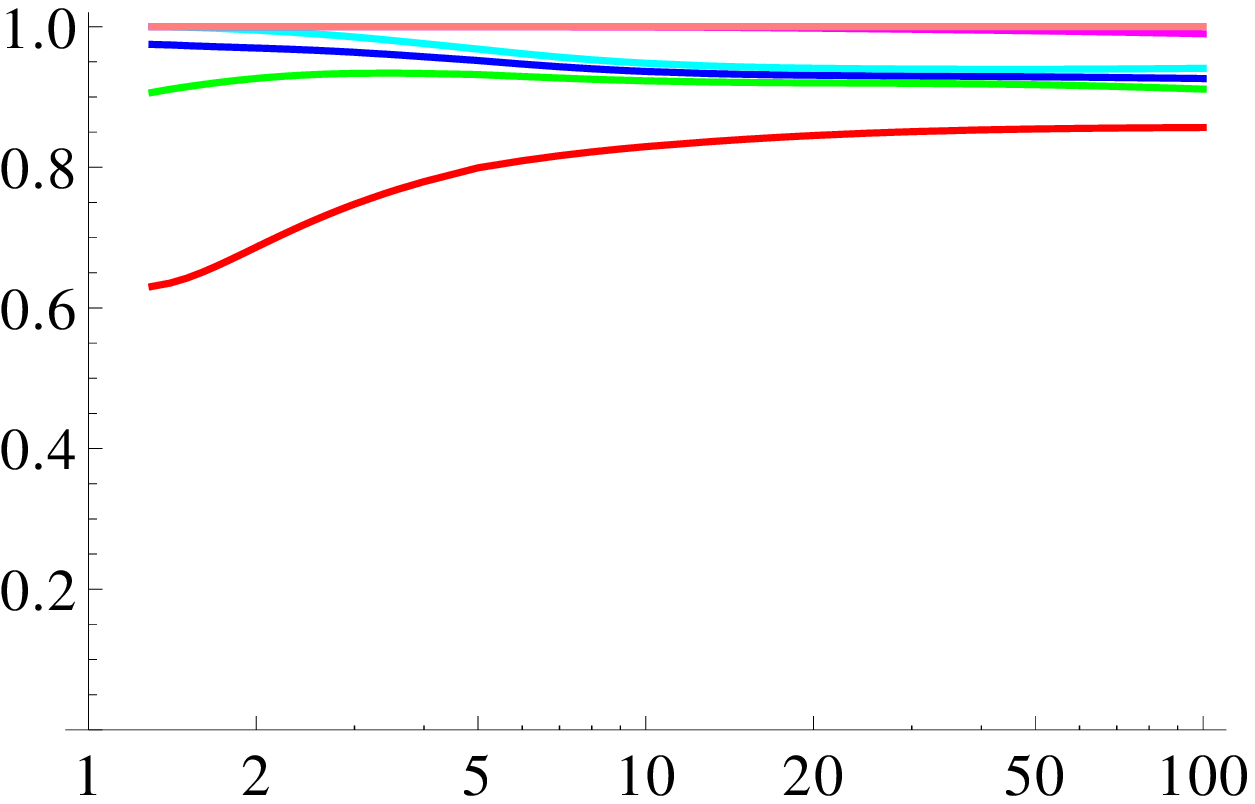}
}
\caption{Fractional flavor decomposition of ``initial-state'' $F_{2,L}^i/F_{2,L}$
vs. $Q$ at N$^3$LO for $x=\{10^{-1},10^{-3},10^{-5}\}$ (left to right)
for $n=2$  scaling.
Reading from the bottom, we plot the cumulative contributions to $F_{2,L}$
from $\{g,u,d,s,c,b\}$, (red, green, blue, cyan, magenta, pink).}
\label{fig:tmp3}
\end{centering}
\end{figure*}

In Figures~\ref{fig:F2trans}  and~\ref{fig:FLtrans} we display
the fractional contributions for the initial-state quarks ($i$)
to the structure functions $F_{2}$ and $F_{L}$,
respectively, for selected $x$ values as a function of
$Q$; here we have used $n=2$ scaling. Reading from the bottom, we
have the cumulative contributions from the $\{g,u,d,s,c,b\}$. 
We observe that for
large $x$ and low $Q$ the heavy flavor contributions are minimal. For
example, for $x=10^{-1}$ we see the contribution of the $u$-quark
comprises $\sim80\%$ of the $F_2$ structure function at low $Q$. In contrast, at $x=10^{-5}$
and large $Q$ we see the $F_2$ contributions of the $u$-quark and $c$-quark
are comparable (as they both couple with a factor 4/9), and the $d$-quark
and $s$-quark are comparable (as they both couple with a factor 1/9).
It is notable that the gluon contribution to $F_{L}$ is significant. 
For  $x=10^{-1}$ this is roughly 40\% throughout the $Q$ range, and can 
be even larger for smaller $x$ values.


%
\begin{figure*}
\begin{centering}
\subfloat[$F_2^j/F_2$ vs. $Q$.
\label{fig:f2RatX135-1}]{
\includegraphics[width=0.32\textwidth]{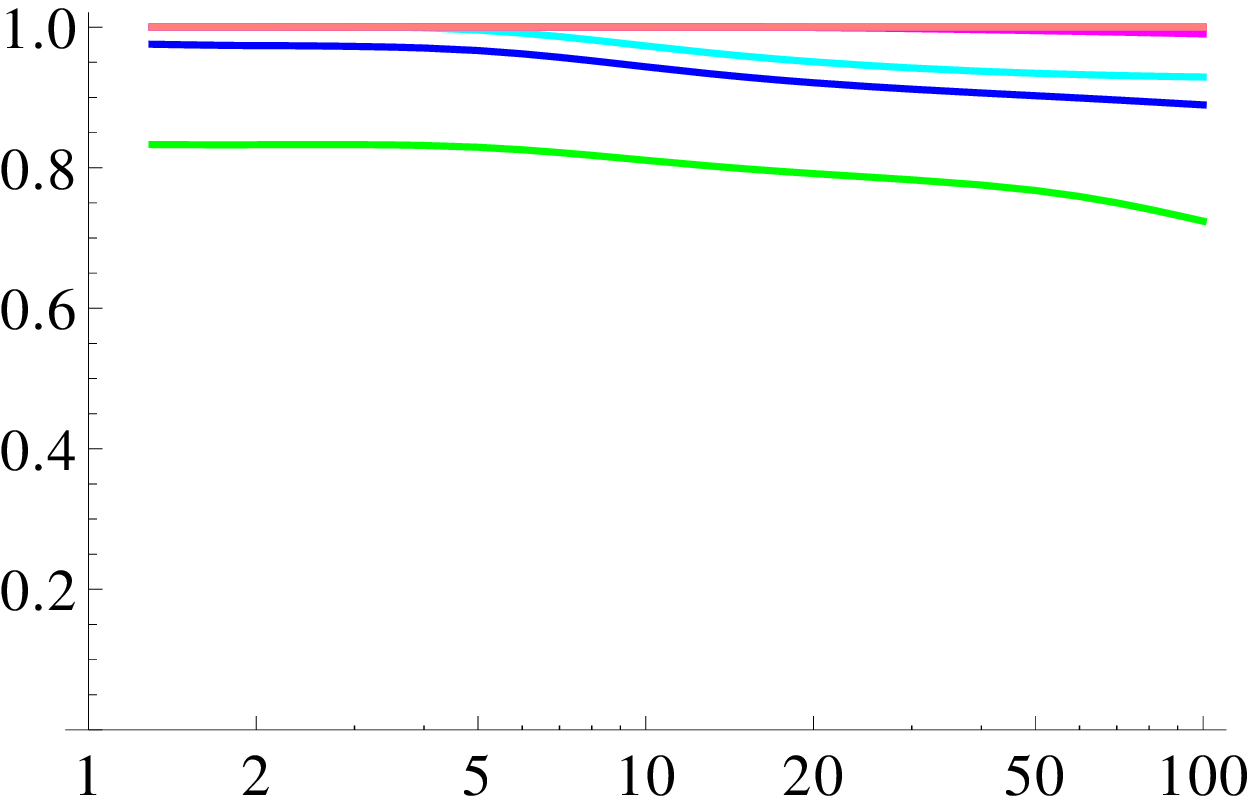}
\hfil
\includegraphics[width=0.32\textwidth]{eps/figN3LO_F2eq11Rat_x3_n2}
\hfil
\includegraphics[width=0.32\textwidth]{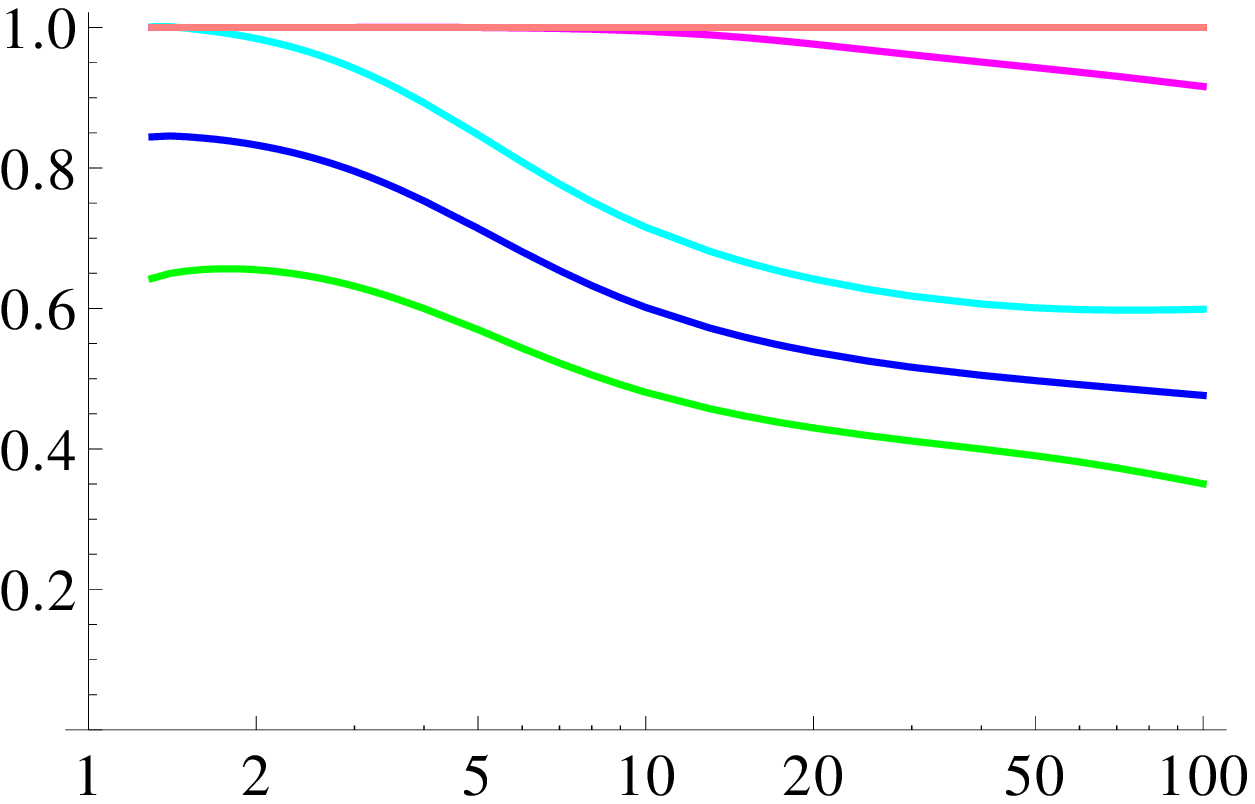}
}

\subfloat[ $F_L^j/F_L$ vs. $Q$.
\label{fig:fLRatX135-1}]{
\includegraphics[width=0.32\textwidth]{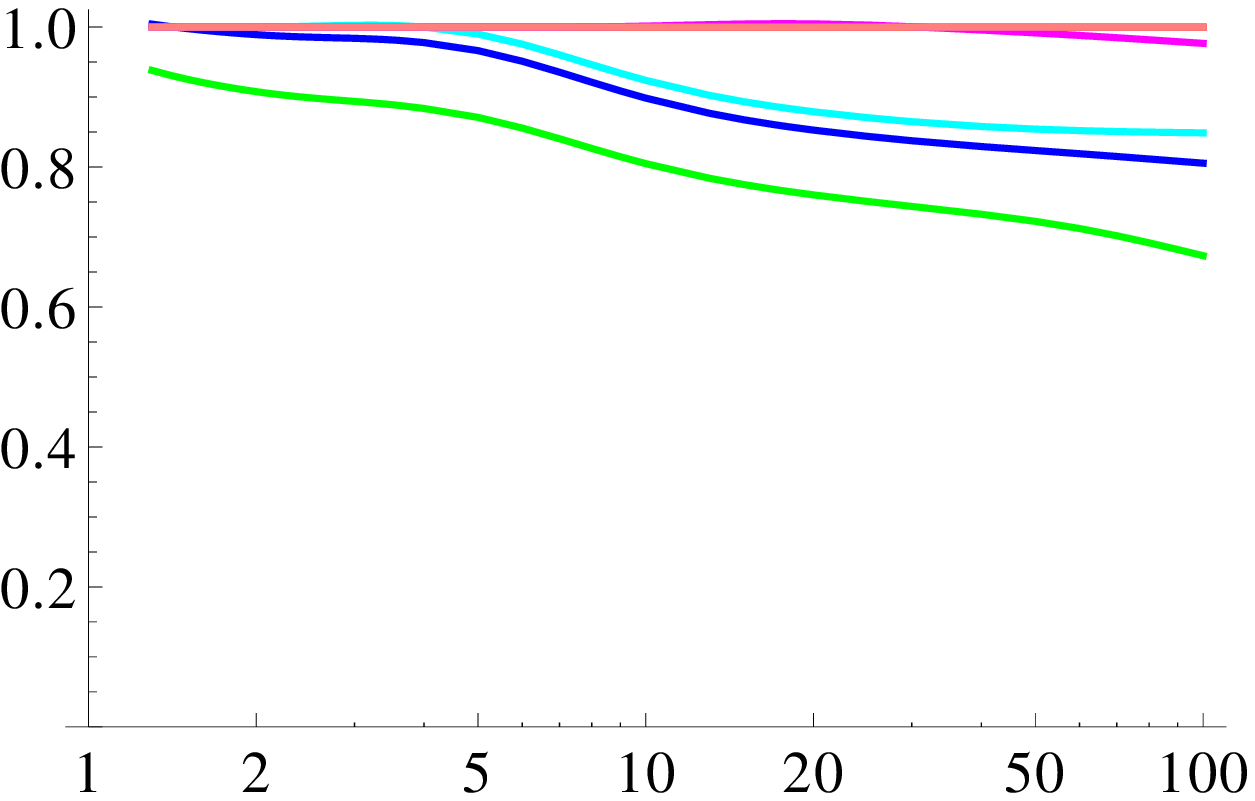}
\hfil
\includegraphics[width=0.32\textwidth]{eps/figN3LO_fLeq11Rat_x3_n2}
\hfil
\includegraphics[width=0.32\textwidth]{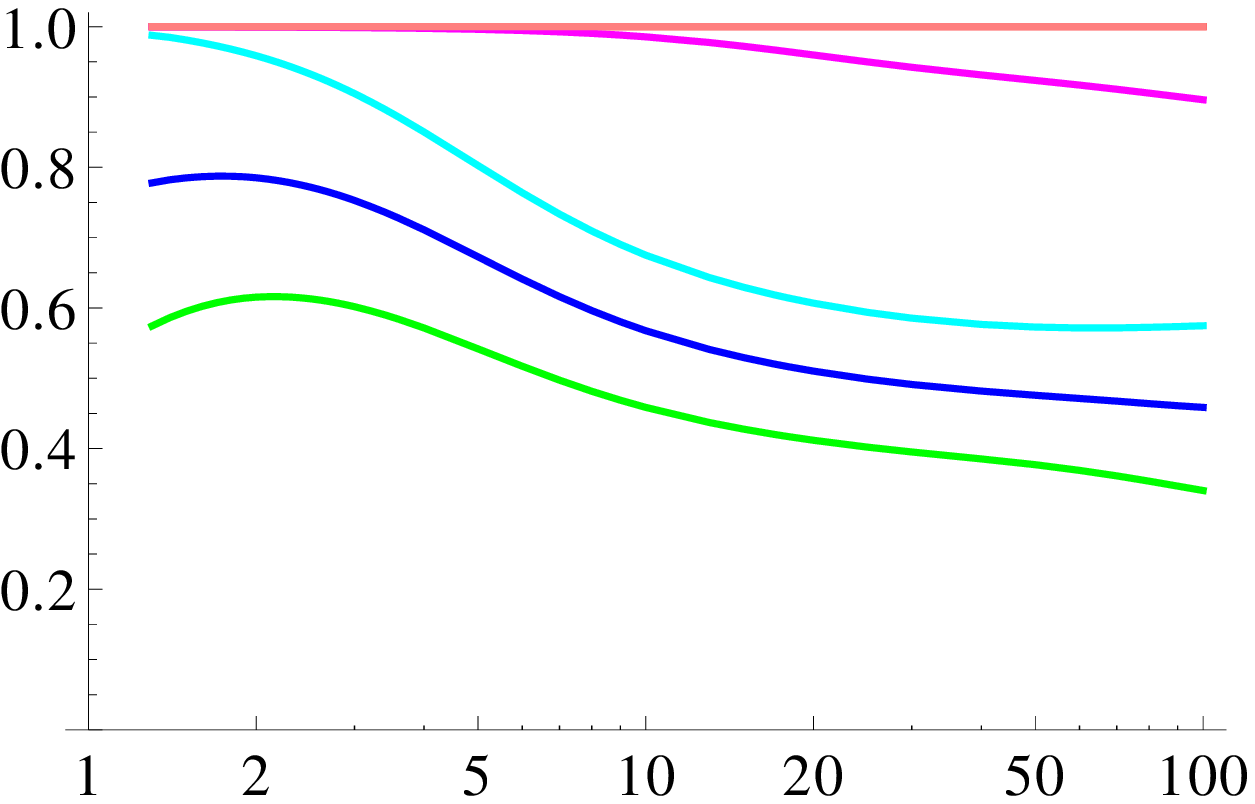}
}
\caption{Fractional contribution for each quark flavor to
$F_{2,L}^j/F_{2,L}$ vs. $Q$ at N$^3$LO for fixed $x=\{10^{-1},10^{-3},10^{-5}\}$
(left to right). Results are displayed for $n=2$ scaling.
Reading from the bottom, we have the cumulative contributions from
the $\{u,d,s,c,b\}$ (green, blue, cyan, magenta, pink).}
\label{fig:eq11}
\end{centering}
\end{figure*}

In Figures \ref{fig:f2RatX135-1} and \ref{fig:fLRatX135-1} we display
the fractional contributions for the final-state quarks ($j$)
 to the structure functions $F_{2}$ and
$F_{L}$, respectively, for selected $x$ values as a function of
$Q$; here we have used $n=2$ scaling. Reading from the bottom, we
have the cumulative contributions from the $\{u,d,s,c,b\}$. 
Again, we observe that for
large $x$ and low $Q$ the heavy flavor contributions are minimal, 
but these can grow quickly as we move to smaller $x$ and larger $Q$.


%
\begin{figure*}
\begin{centering}
\subfloat[$F_{2}$ vs. $Q$.
\label{fig:f2orders}]{
\includegraphics[width=0.32\textwidth]{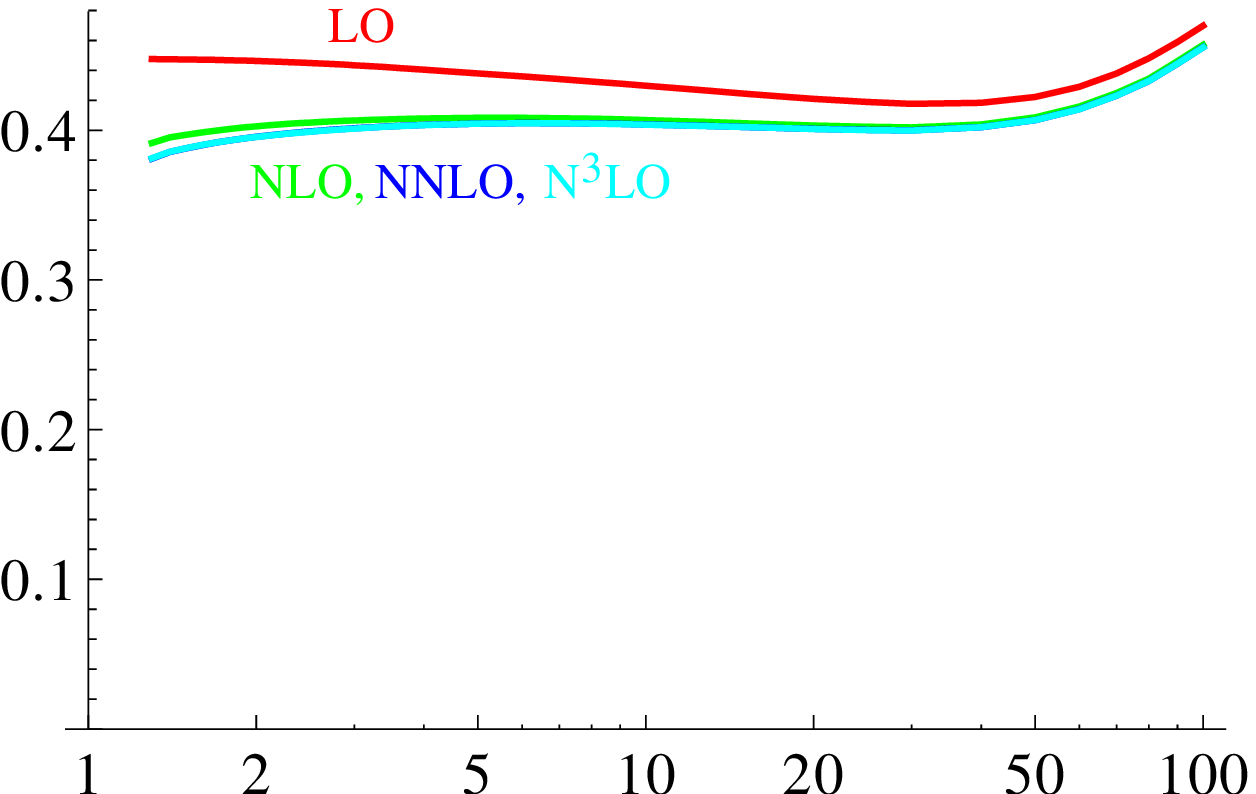}
\hfil
\includegraphics[width=0.32\textwidth]{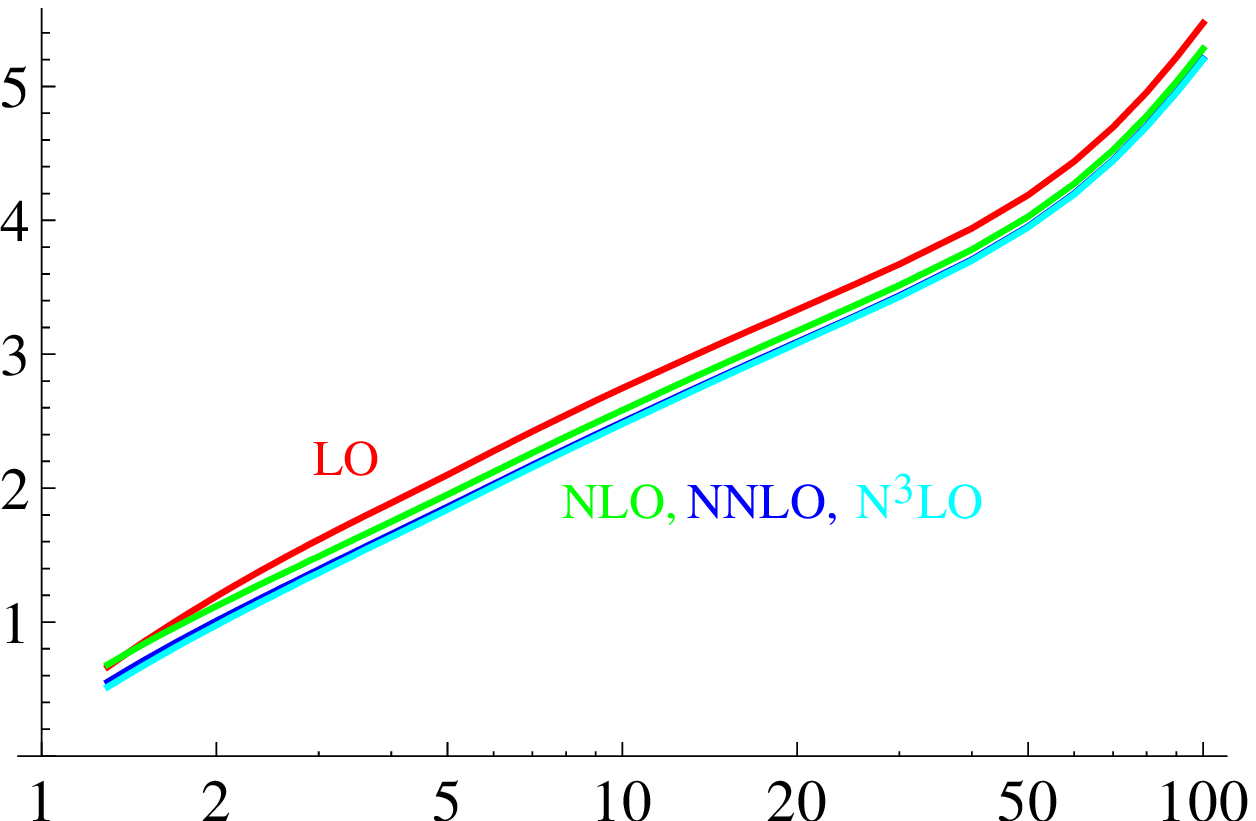}
\hfil
\includegraphics[width=0.32\textwidth]{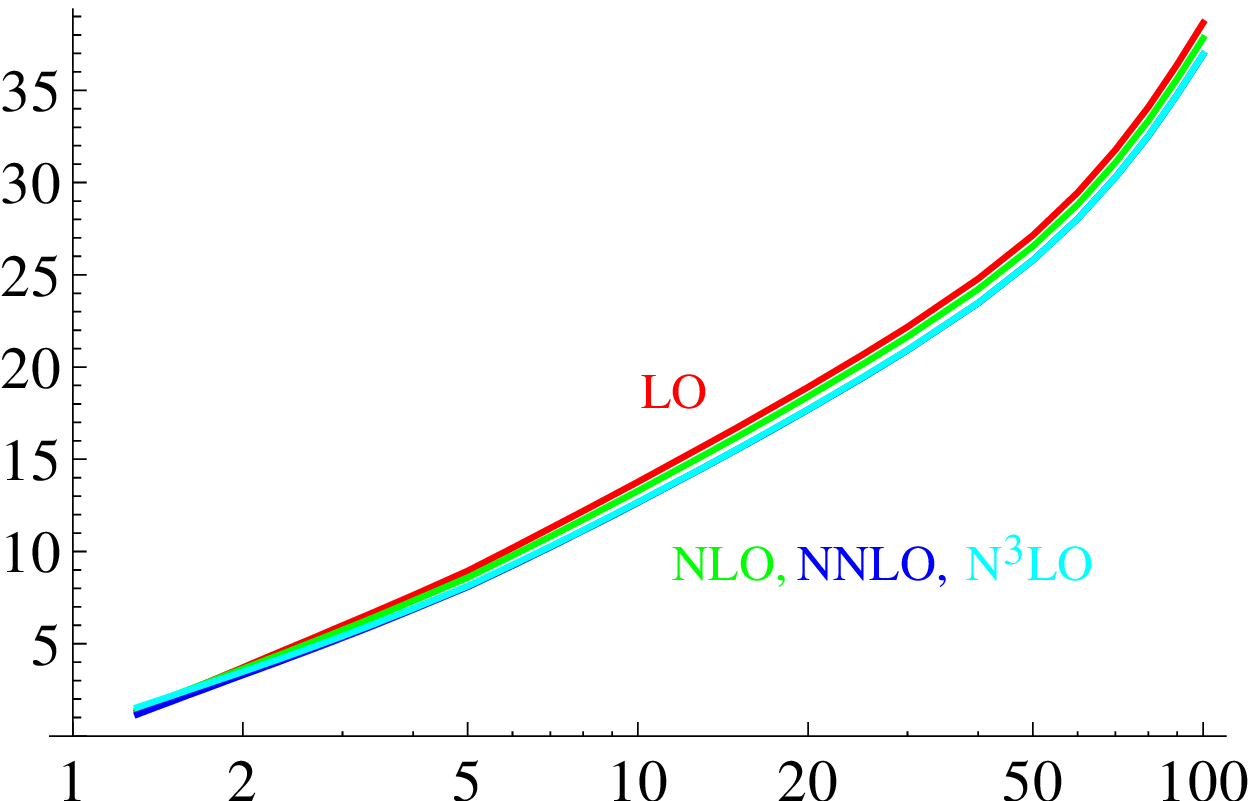}
}

\subfloat[$F_{L}$ vs. $Q$.
\label{fig:fLorders}]{
\includegraphics[width=0.32\textwidth]{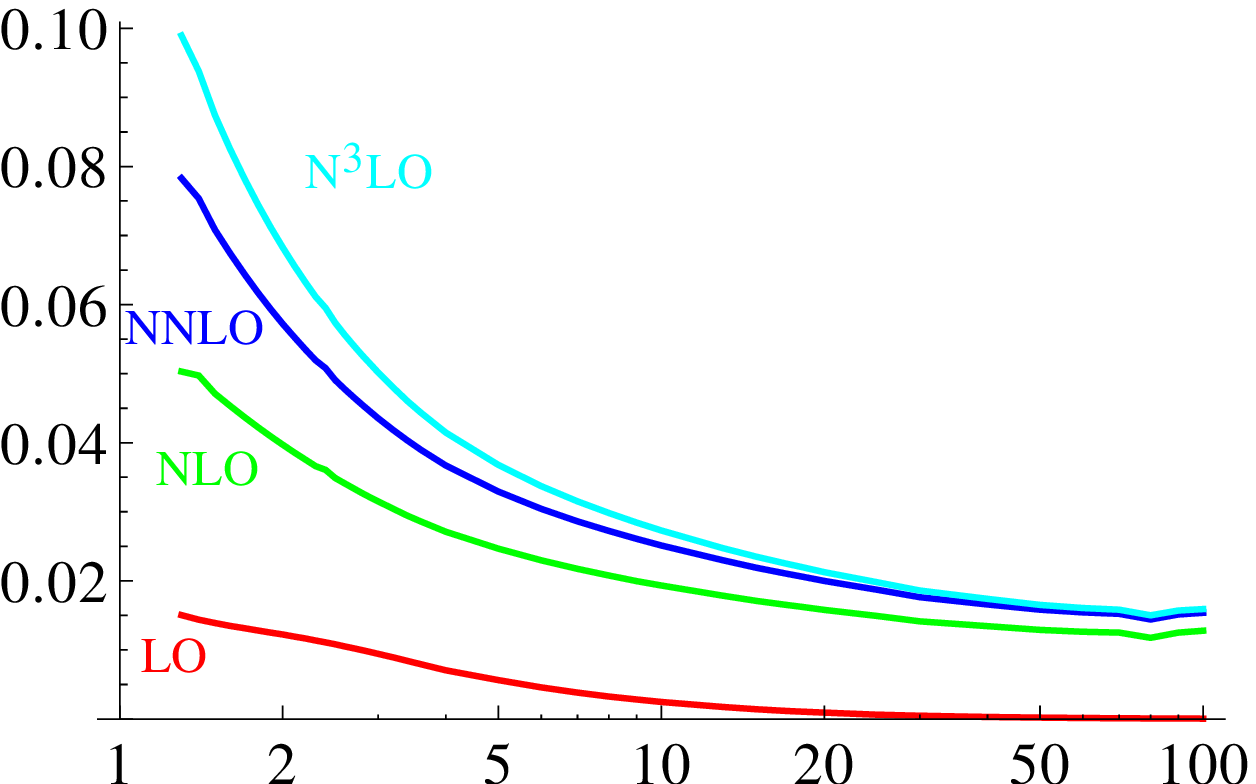}
\hfil
\includegraphics[width=0.32\textwidth]{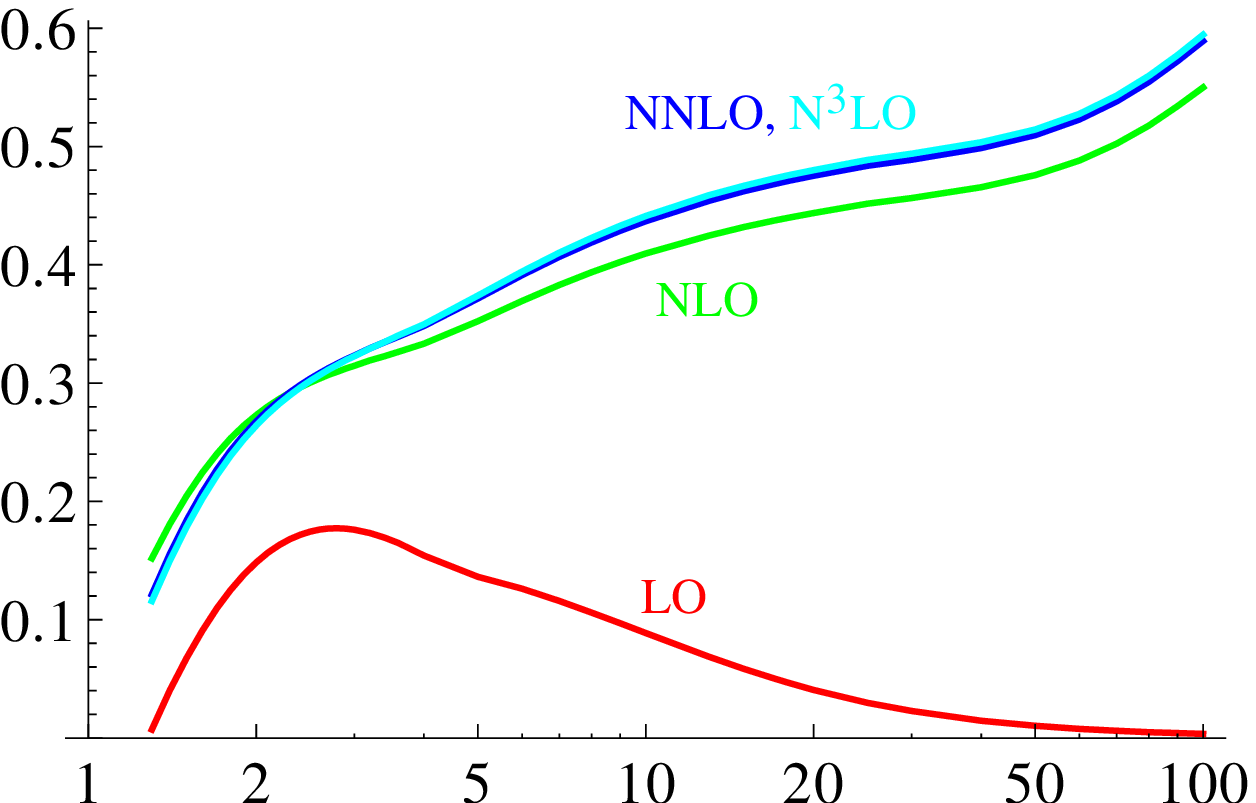}
\hfil
\includegraphics[width=0.32\textwidth]{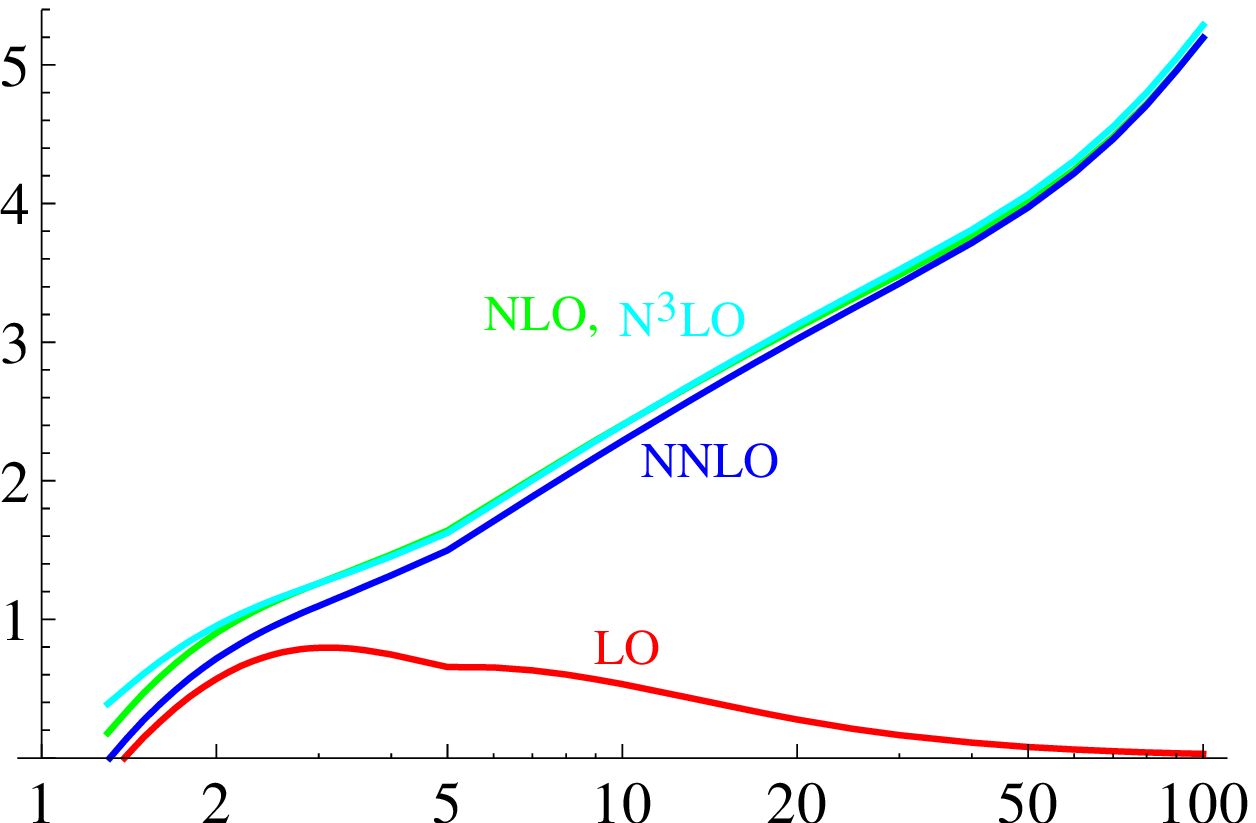}
}

\caption{$F_{2,L}$ vs. $Q$ at \{LO, NLO, NNLO, N$^3$LO\}
(red, green, blue, cyan) for fixed $x=\{10^{-1},10^{-3},10^{-5}\}$
(left to right) for $n=2$ scaling.}
\label{fig:F2L_orders}
\end{centering}
\end{figure*}

In Figure~\ref{fig:f2orders} we display the results for $F_{2}$
vs. $Q$ computed at various orders. For large $x$ (c.f. $x=0.1$) we
find the perturbative calculation is particularly stable; we see
that the LO result is within 20\% of the others at small $Q$, and
within 5\% at large $Q$. The NLO is within 2\% at small $Q$, and
indistinguishable from the NNLO and N$^3$LO for $Q$ values above $\sim10$~GeV.
The NNLO and N$^3$LO results are essentially identical throughout the
kinematic range. For smaller $x$ values ($10^{-3}$, $10^{-5}$) the
contribution of the higher order terms increases. Here, the
NNLO and N$^3$LO coincide for $Q$ values above $\sim5$~GeV, but the NLO
result can differ by $\sim5\%$.

In Figure~\ref{fig:fLorders} we display the results for $F_{L}$
vs. $Q$ computed at various orders. In contrast to $F_{2}$,
we find the NLO corrections are large for $F_L$; this is because the LO
$F_{L}$ contribution (which violates the Callan-Gross relation) is suppressed
by $(m^{2}/Q^{2})$ compared to the dominant gluon contributions which
enter at NLO. Consequently, we observe (as expected) that the LO result
for $F_{L}$ receives large contributions from the higher order terms.
Essentially, the NLO is the first non-trivial order for $F_{L}$, and
the subsequent contributions then converge. For example, at large
$x$ (c.f. $x=0.1$) for $Q\sim10$~GeV we find the NLO result yields
$\sim60$ to $80\%$ of the total, the NNLO is a $\sim20\%$ correction,
and the N$^3$LO is a $\sim10\%$ correction. For lower $x$ values ($10^{-3}$,
$10^{-5}$) the convergence of the perturbative series improves, and
the NLO results is within $\sim10\%$ of the N$^3$LO result. Curiously,
for $x=10^{-5}$ the NNLO and N$^3$LO roughly compensate each other so
that the NLO and the N$^3$LO match quite closely for $Q\geq2$~GeV.

While the calculation of $F_L$ is certainly more challenging,
examining Fig.~\ref{fig:slac-4-4} we see that for most of the relevant
kinematic range probed by HERA the theoretical calculation is quite
stable.
For example, in the high $Q^2$ region where HERA is probing
intermediate $x$ values ($x\sim 10^{-3}$) the spread of the $\chi(n)$
scalings is small. The challenge arises in the low $Q$ region ($Q\sim
2$~GeV) where the $x$ values are $\sim 10^{-4}$; in this region, there
is some spread between the various curves at the lowest $x$ value
$(\sim 10^{-5})$, but for $x\sim 10^{-3}$ this is greatly reduced.

\section{Conclusions\label{sec:conclusion} }

We extended the ACOT calculation for DIS structure functions 
to  N$^3$LO by combining  the exact ACOT
scheme at NLO with a $\chi(n)$-rescaling which allows us to
include the leading mass dependence at NNLO and N$^3$LO.
Using the full ACOT calculation at NLO, we demonstrated that the heavy
quarks mass dependence for the DIS structure functions is dominated by
the kinematic mass contributions, and this can be implemented via a
generalized $\chi(n)$-rescaling prescription.

We studied the $F_2$ and $F_L$ structure functions as a function of
$x$ and $Q$.  We examined the flavor decomposition of these structure
functions, and verified that the heavy quarks were appropriately
suppressed in the low $Q$ region.
We found the results for $F_2$ were very stable across the full kinematic range
for $\{x,Q\}$, and the contributions from the NNLO and N${}^3$LO terms
were small.
For $F_L$, the higher order terms gave a proportionally larger
contribution (due to the suppression of the LO term from the
Callan-Gross relation); nevertheless, the contributions from the NNLO
and N${}^3$LO terms were generally small in the region probed by HERA.

The result of this calculation was to obtain precise predictions for
the inclusive $F_2$ and $F_L$ structure functions which can be used
to analyze the HERA data.

\section*{Acknowledgment}

We thank 
M.~Botje, 
A.~M.~Cooper-Sarkar, 
A.~Glazov,
C.~Keppel, 
J.~G. Morf\'in, 
P.~Nadolsky, 
M.~Guzzi,
J.~F.~Owens,
V.~A.~Radescu,
and
A.~Vogt
for discussions.
F.I.O., I.S., and J.Y.Y.\ acknowledge the hospitality of 
CERN, DESY, Fermilab, and Les Houches where a portion of this work
was performed. This work was partially supported by the U.S.\ Department
of Energy under grant DE-FG02-04ER41299, and the Lightner-Sams Foundation.
F.I.O thanks the Galileo
Galilei Institute for Theoretical Physics for their hospitality
and the INFN for partial support during the
completion of this work.
The research of T.S. is supported by a fellowship from the
Th\'eorie LHC France initiative funded by the CNRS/IN2P3.
This work has been supported by  {\it Projet international de cooperation
scientifique} PICS05854 between France and the USA.
The work of J.~Y.~Yu was supported by the Deutsche Forschungsgemeinschaft
(DFG) through grant No.~YU~118/1-1.

\newpage
\bibliographystyle{utphys_spires}
\bibliography{bibEpiphany}

\end{document}